\documentclass[12pt,twoside]{article}
\usepackage[mathscr]{eucal}
\usepackage{tocloft}
\usepackage[nosort]{cite}
\usepackage{amsmath,amsfonts,amssymb,amsthm,ulem,latexsym}
\usepackage[dvips]{graphicx}
\usepackage{tabularx}
\usepackage{times}
\usepackage{multibox}

\makeatletter

\renewcommand\section{\@startsection{section}{1}{\z@}
                                   {-3.5ex \@plus -1ex \@minus -.2ex}
                                   {2.3ex \@plus .2ex}
                                   {\normalfont\large\bfseries}}
\renewcommand\subsection{\@startsection{subsection}{2}{\z@} 
                                   {-3.25ex\@plus -1ex \@minus -.2ex}
                                   {1.5ex \@plus .2ex}
                                   {\normalfont\normalsize\bfseries}}
\renewcommand\subsubsection{\@startsection{subsubsection}{3}{\z@}
                                   {-3.25ex\@plus -1ex \@minus -.2ex}
                                   {1.5ex \@plus .2ex}
                                   {\normalfont\normalsize\bfseries}}
\renewcommand\paragraph{\@startsection{paragraph}{4}{\z@}
                                   {3.25ex \@plus1ex \@minus.2ex}
                                   {-1em}
                                   {\normalfont\normalsize\bfseries}}

\makeatother

\advance\textheight\topskip \textwidth 35.5pc \oddsidemargin 20pt
\renewcommand{\tilde}{\widetilde}
\renewcommand{\hat}{\widehat}

\newcommand{\SVir}{\mathrm{SVir}_2}
\def\cM{\mathcal{M}}
\def\cN{\mathcal{N}}

\def\cW{\mathcal{W}}

\numberwithin{equation}{section} \makeatletter
\@addtoreset{equation}{section}

\makeatletter
\def\Appendix{\appendix
  \def\@seccntformat##1{Appendix~\csname the##1\endcsname.~~}}
\makeatother
\newcolumntype{x}[1]{%
>{\centering\hspace{0pt}}m{#1}}%
\newcolumntype{w}[1]{%
>{\raggedright\hspace{0pt}}m{#1}}%
\newcolumntype{z}[1]{%
>{\raggedleft\hspace{0pt}}m{#1}}%

\newdimen\tableauside\tableauside=1.0ex
\newdimen\tableaurule\tableaurule=0.4pt
\newdimen\tableaustep
\def\phantomhrule#1{\hbox{\vbox to0pt{\hrule height\tableaurule
width#1\vss}}}
\def\phantomvrule#1{\vbox{\hbox to0pt{\vrule width\tableaurule
height#1\hss}}}
\def\sqr{\vbox{%
  \phantomhrule\tableaustep

\hbox{\phantomvrule\tableaustep\kern\tableaustep\phantomvrule\tableaustep}%
  \hbox{\vbox{\phantomhrule\tableauside}\kern-\tableaurule}}}
\def\squares#1{\hbox{\count0=#1\noindent\loop\sqr
  \advance\count0 by-1 \ifnum\count0>0\repeat}}
\def\tableau#1{\vcenter{\offinterlineskip
  \tableaustep=\tableauside\advance\tableaustep by-\tableaurule
  \kern\normallineskip\hbox
    {\kern\normallineskip\vbox
      {\gettableau#1 0 }%
     \kern\normallineskip\kern\tableaurule}%
  \kern\normallineskip\kern\tableaurule}}
\def\gettableau#1 {\ifnum#1=0\let\next=\null\else
  \squares{#1}\let\next=\gettableau\fi\next}

\newcommand{\be}{\begin{equation}}
\newcommand{\ee}{\end{equation}}
\newcommand{\bea}{\begin{eqnarray}}
\newcommand{\eea}{\end{eqnarray}}
\newcommand{\ba}{\begin{array}}
\newcommand{\ea}{\end{array}}
\newcommand{\id}{\hbox{1\kern-.27em l}}

\newcommand{\ZZ}{\mathbb{Z}}
\newcommand{\CC}{\mathbb{C}}
\newcommand{\RR}{\mathbb{R}}

\newcommand{\half}{ {\textstyle \frac{1}{2}  } }
\newcommand{\al}{\alpha}
\newcommand{\ga}{\gamma}

\newcommand{\bet}{\beta}

\newcommand{\ka}{\kappa}

\newcommand{\ep}{\epsilon}

\newcommand{\la}{\lambda}

\newcommand{\om}{\omega}
\newcommand{\Om}{\Omega}

\newcommand{\De}{\Delta}

\newcommand{\Ups}{\Upsilon}
\newcommand{\pa}{\partial}
\newcommand{\rar}{\rightarrow}

\newcommand{\non}{\nonumber}
\newcommand{\lb}{\langle}
\newcommand{\rb}{\rangle}

\newcommand{\tia}{\tilde{a}}

\newcommand{\tY}{\tilde{Y}}

\newcommand{\SU}{\mathrm{SU}}

\newcommand{\SL}{\mathrm{SL}}
\newcommand{\GL}{\mathrm{GL}}

\newcommand{\sll}{\mathrm{sl}}
\newcommand{\U}{\mathrm{U}}
\newcommand{\ul}{\mathrm{u}}

\newcommand{\ts}{\textstyle}
\newcommand{\ds}{\displaystyle}
\begin{document}

\begin{flushright}
FIAN-TD-2012-09 
\end{flushright}
\vspace{2mm}

\begin{center}

{\large\sf {$\cN\!=\!2$} superconformal blocks and  instanton partition functions}

\vspace*{6mm}
{V. Belavin$^{1}$ and Niclas Wyllard$^{2}$}

\vspace*{4mm}
$^1$~\parbox[t]{0.85\textwidth}{\normalsize\raggedright
Theoretical Department, Lebedev Physical Institute, RAS, Moscow, Russia}

$\,^2$ ~\parbox[t]{0.85\textwidth}{\normalsize\raggedright
{\tt \small n.wyllard@gmail.com}}

\vspace{0mm}

\begin{abstract}\vspace*{1pt}
We consider the problem of computing (irregular) conformal blocks in $2d$ CFTs whose chiral symmetry algebra is the $\cN\!=\!2$ superconformal algebra.

Our construction uses two ingredients:~(i) the relation between the representation theories of the $\cN\!=\!2$ superconformal algebra and the affine $\widehat{\sll}(2)$ algebra, extended to the level of the conformal blocks, and (ii) the relation between $\widehat{\sll}(2)$ conformal blocks and  instanton partition functions in the $4d$ $\cN\!=\!2$ $\SU(2)$ gauge theory with a surface defect. 
By combining these two facts we derive combinatorial expressions for the $\cN\!=\!2$ superconformal blocks in the Gaiotto limit.
\end{abstract}
\end{center}

\vspace*{-2mm}

\setcounter{tocdepth}{1}
{ \small \tableofcontents }

\newpage

\section{Introduction}
Conformal field theories in two dimensions with $\cN\!=\!(2,2)$ supersymmetry 
have been intensely studied over the years. The chiral symmetry algebra of such theories is the $\cN=2$ superconformal algebra \cite{Ademollo:1975}. The full superconformal field theory contains two commuting copies (holomorphic and antiholomorphic) of the $\cN\!=\!2$ superconformal  algebra. The main reason why $\cN\!=\!(2,2)$ supersymmetry is interesting derives from its  applications in  superstring theory and in topological field theories.
However, there are still important aspects of $\cN\!=\!(2,2)$ superconformal theories that have not been investigated. In particular, the explicit construction of conformal blocks is missing. 

Recently, following \cite{Alday:2009a}, several relations between different types of $2d$ conformal field theories and instanton partition functions and their associated moduli spaces in $4d$ $\cN\!=\!2$ supersymmetric gauge theories have been proposed. This type of correspondence (AGT) has turned out to be an efficient tool for constructing conformal blocks. A natural question to ask is: what is the gauge theory and the instanton moduli space corresponding to the $\cN\!=\!2$ superconformal algebra? Even though there has been a lot of progress in understanding AGT-type relations, the general framework which would allow one to directly answer this question remains rather obscure.

Last year, it was conjectured~\cite{Belavin:2011a} that there should be a relation between $\cN\!=\!2$ $\SU(N)$ 
gauge theories on $\mathbb{R}^4/\mathbb{Z}_p$ and the coset conformal field theories 
\begin{equation}\label{BF-coset}
\widehat{\ul}(1)  
\! \times \widehat{\sll}(p)_N{\times} \frac{ \widehat{\sll}(N)_{\ka}  \! \times \widehat{\sll}(N)_p}{\widehat{\sll}(N)_{\ka+p}}\,,
\end{equation}
where the parameter $\ka$ is related to the equivariant deformation parameters \cite{Moore:1997,Nekrasov:2002} in the gauge theory. 
A deeper understanding of the proposal in \cite{Belavin:2011a} as well as various generalisations  have been achieved in~\cite{Nishioka:2011, Bonelli:2011a,Belavin:2011b, Bonelli:2011b,Wyllard:2011, Ito:2011, Alfimov:2011,Belavin:2011c}.
Among the theories of type \eqref{BF-coset} one finds various extensions of the Virasoro algebra. For example, in the $\SU(2)$ case $p\!=\!2$ corresponds to the $\cN\!=\!1$ superconformal algebra (the Neveu-Schwarz-Ramond algebra) and $p\!=\!4$ corresponds to the abelianly braided spin $4/3$ parafermionic algebra \cite{Fateev:1985a}. 
On the other hand, the $\cN\!=\!2$ superconformal algebra does not fit into this general scheme. 

The aim of the present paper is to study the $\cN\!=\!2$ superconformal algebra  from the point of view
of the relations between $2d$ conformal field theories and instanton moduli spaces. 
Our starting point is the realisation \cite{DiVecchia:1986} of the $\cN\!=\!2$ superconformal algebra in terms of the coset 
\be \label{N2coset}
\frac{\widehat{\sll}(2)_k\! \times\widehat{\ul}(1)}{\widehat{\ul}(1)} \,,
\ee 
where the parameter $k$ is the level of the affine $\widehat{\sll}(2)$ algebra. 
The coset construction (\ref{N2coset}) implies that there is a close relation between the representation theory of the $\cN\!=\!2$ superconformal algebra and the representation theory of  the affine $\widehat{\sll}(2)$ algebra \cite{Feigin:1997}.  A priori this is quite surprising since the two algebras as well as their highest weight representations look completely different. However, the relation does not involve a map between the standard highest weight representations of the two algebras. Instead ``standard" representations in one algebra are related to ``non-standard" representations in the other algebra. In particular, the standard (massive) highest weight module of the $\cN\!=\!2$ algebra is related to a non-standard module 
 --- the so called {\it relaxed} module --- of the affine $\widehat{\sll}(2)$ algebra. 

In this paper we extend the relation in \cite{Feigin:1997} to the level of the conformal blocks. This allows us to express $\cN\!=\!2$  superconformal blocks in terms of $\widehat{\sll}(2)$ 
conformal blocks for primary fields associated with the relaxed $\widehat{\sll}(2)$ representations. In turn we  observe that the relaxed $\widehat{\sll}(2)$ conformal blocks can be obtained from the ordinary (unrelaxed) $\widehat{\sll}(2)$ conformal blocks via a certain well defined analytic continuation that we explictly describe. 
Unrelaxed $\widehat{\sll}(2)$ conformal blocks have been shown to be equal to instanton partition functions in $4d$ $\cN\!=\!2$ $\SU(2)$ gauge theories with a surface defect (operator) \cite{Braverman:2004a,Alday:2010}.  By implementing the analytic continuation in the instanton partition function we obtain an explicit combinatorial expression for  $\cN\!=\!2$  superconformal blocks. In the present paper we
restrict our attention to the case of irregular superconformal blocks, i.e.~the four-point block on the sphere in the Gaiotto (Whittaker) limit. 

This paper is organised as follows. In the next section we review the $\cN\!=\!2$ superconformal algebra and its highest weight representations and introduce a Whittaker state. In section \ref{ssl2} we review the $\widehat{\sll}(2)$ algebra and its representations, focusing on the relaxed modules. We also introduce a Whittaker state in the relaxed module. Then in section \ref{sGai} we describe how the $\cN\!=\!2$ Whittaker state arises from a limit of the four-point conformal blocks on the sphere.  The interrelations between the $\cN\!=\!2$, relaxed $\widehat{\sll}(2)$, and ordinary $\widehat{\sll}(2)$ modules is discussed in section \ref{sRel}, where we also extend the relations to the level of the conformal blocks. In section \ref{sSurf} we discuss the instanton partition functions in $\SU(N)$ gauge theories with a surface defect and in particular write the instanton partition function as a sum over coloured Young diagrams. Our main result is contained in section \ref{sComb} where we put all the pieces together and derive a combinatorial expression for the norm of the $\cN\!=\!2$ Whittaker state. Finally in section \ref{Concl} we present our conclusions and discuss some possible extensions of our work. 
The appendices contain more technical aspects of our derivations.

%%%%%%%%%%%%%%%%%%%%%%%%%%%%%%%%%
\section{The $\cN\!=\!2$ superconformal algebra} \label{sN2}
%%%%%%%%%%%%%%%%%%%%%%%%%%%%%%%%%

In this section we recall some details about the $\cN\!=\!2$ superconformal algebra and its representations. We also define a Whittaker state and compute its norm for some low levels.

%%%%%%%%%%%%%%%%%%%%%%%%%%%%%%%%
\subsection{The $\cN\!=\!2$ algebra and its modules} 
%%%%%%%%%%%%%%%%%%%%%%%%%%%%%%%%

The $\cN\!=\!2$ superconformal algebra (which we sometimes denote $\SVir$) is generated by
 the stress-energy tensor $T(z)$,  two dimension $3/2$ Grassmann-odd  currents $G^+(z)$, $G^-(z)$,   and  the dimension one current $H(z)$. In terms of modes the 
$\cN\!=\!2$ superconformal algebra satisfies the following comutation relations
\begin{eqnarray}
\label{N2alg}
&&[L_m, L_n] = (m-n) L_{n+m} + \frac{c}{4}(m^3-m)\delta_{m+n,0}\;,
\nonumber\\
&&\left[L_m, H_n\right] = -n H_{m+n}\;,\,\,\,\,\,\,\left[L_m, G_r^{\pm}\right]= 
(\frac{m}{2}-r)G_{m+r}^{\pm}\;,
\nonumber\\
&&\left[H_m, H_n\right] = c\; m\; \delta_{m+n,0}\;,\,\, [H_m, G_r^{\pm}] = \pm G_{m+r}^{\pm}\;,
\\
&&\{G_r^+, G_s^-\} =2 L_{r+s}+(r-s) H_{r+s}+c(r^2-{\ts \frac 14 })\delta_{r+s,0}\;.\nonumber
\end{eqnarray}
Here $c$ is the central charge. In what follows we consider only the NS sector. The NS sector is closed under the operator product expansion and is defined by
\begin{equation}
m,n \in \mathbb{Z}\,\,\,\,\text{and}\,\,\,\,r,s \in \mathbb{Z} +\half\;.
\end{equation}
The hermitean conjugates of the modes are $L_n^\dag = L_{-n}$, $H_n^\dag = H_{-n}$, and  $(G_r^-)^\dag = G^+_{-r}$. 

We denote the highest weight state corresponding to the primary field $V_{\Delta}^\omega(z)$ of the $\cN=2$ algebra by $ |\De,\om\rb $ ($\Delta,\omega\in\CC$). It satisfies 
\be
L_0 |\De,\om\rb =\Delta| \De,\om\rb \,,\qquad H_0 |\De,\om\rb =\omega |\De,\om\rb \,,
\ee
and is such that the subalgebra $\mathrm{SVir}_2^+$ acts trivially:
\be
L_{n>0} |\De,\om\rb=0 \,,\qquad H_{n>0}  |\De,\om\rb=0 \,,\qquad G^{\pm}_{r>0}  |\De,\om\rb=0\,. 
\ee
The Verma module $M_{\Delta,\omega}$  is spanned by the descendants obtained by acting on the highest weight state with  negative (creation) modes; in other words, it is given by the action of the enveloping algebra of $\mathrm{SVir}_2^-$ acting on the highest weight state:
\begin{equation}
M_{\Delta,\omega}=\mathfrak{U}(\mathrm{SVir}_2^-)  |\De,\om\rb.
\end{equation}
The Verma module has the $L_0$-level and $H_0$-charge decomposition: 
\begin{align}
M_{\Delta,\omega}=\bigoplus_{n\in\mathbb{Z}}\bigoplus_{N\in n^2/2+\mathbb{Z}_+}
 M_{\Delta,\omega}^{N,n},
\end{align}
where
\begin{align}  
M_{\Delta,\omega}^{N,n}=\{ |v\rb \in M_{\Delta,\omega} \mid L_0 |v\rb =(\Delta+N)| v\rb \,,\,\, H_0 |v\rb =(\omega+n)|v\rb\}.
\end{align}
The matrix of inner products of descendants (the Shapovalov matrix, $\mathcal{S}$) is block-diagonal in this decomposition.

%%%%%%%%%%%%%%%%%%%%%%%%%%%%%%%%%%%%%%%
\subsection{A Whittaker state of the $\cN\!=\!2$ superconformal algebra} 
%%%%%%%%%%%%%%%%%%%%%%%%%%%%%%%%%%%%%%%

A Whittaker state (vector) is a special state in a Verma module that has certain eigenvalue properties and is not annihilated by all the positive modes. A priori there are several possible Whittaker states that one can define. A natural defintion of a Whittaker state  arises in a limiting case  of a four-point conformal block \cite{Gaiotto:2009b}, leading to a so called irregular conformal block which is equal to the norm of the Whittaker state. 
As we will show in section \ref{sGai}, in this limit  the four-point $\cN\!=\!2$ block splits into two completely decoupled sectors (``BPS" and ``anti-BPS"), where each sector has an associated  Whittaker state. Since the two decoupled  sectors and their Whittaker states are essentially equivalent we may without loss of generality focus on only one sector.

We define the $\cN\!=\!2$ Whittaker state by the following equations\footnote{The definition of the Whittaker state in the other sector is obtained by interchanging $+\leftrightarrow -$.}
\be \label{Whit}
G_{1/2}^- |W\rb = \sqrt{z}\, |W\rb^- , \quad G_{1/2}^+ |W\rb^- = 2 \sqrt{z} \, |W\rb \quad \Rightarrow \;\, L_{1}|W\rb = z |W\rb\,,
\ee
where $H_1$ and all other positive modes annihilate  $|W\rb$ and $|W\rb^-$ (note that $G^{\pm}_{1/2}$ and $H_1$ generate $\mathrm{SVir}_2^+$). The Whittaker state can be decomposed as
\begin{equation} \label{Whitexp}
|W\rb =\sum_{N=0}^{\infty}  z^N |N\rangle\,, \qquad |W\rb^- =\sum_{N\in \ZZ^+ -\frac{1}{2}} \!\!z^{N} |N\rangle^- \,,
\end{equation}
and we call the states $|N\rangle$, $|N\rangle^-$  Gaiotto states. 

Just like for the Virasoro case \cite{Gaiotto:2009b,Marshakov:2009}, it is easy to see that the norm of the Whittaker state at a given level is equal to a particular diagonal element of the inverse of the Shapovalov matrix (here ${}^-\langle N|$ denotes the hermitean conjugate of $N\rangle^-$): 
\begin{eqnarray} \label{Whitnorm}
&&\,\,\, \, \langle N| N\rangle \,\,\,= \, \mathcal{S}^{-1}(L_{-1}^N;L_{-1}^N)\,,
\;\quad\qquad\qquad \qquad \text{for  }N\text{ integer},\\
&&{}^-\langle N| N\rangle^- = 2^2\, \mathcal{S}^{-1}(L_{-1}^{N-\frac 12} G_{-\frac 12}^{-} ;L_{-1}^{N-\frac 12} G_{-\frac 12}^{-}) \,,
\quad \text{for }N\text{ half-integer}.
\end{eqnarray}
The Gaiotto states satisfy the recursion  relations (see section \ref{sGai} for details)
\begin{eqnarray}\label{chainN2}
&&G_{\frac 12}^{-}|N\rangle\,=\,\,|N{-}\half \rangle^-, \, \nonumber\\
&&G_{\frac 12}^{+}|N{+}\half \rangle^-=\,2 |N\rangle \,.
\end{eqnarray}
Note that when combined with (\ref{Whitexp}) the recursion relations (\ref{chainN2}) imply (\ref{Whit}).  
The norm of the Whittaker state at a given level can be obtained from (\ref{Whitnorm}).

We close this section with some examples. At levels $1$ and $3/2$ the standard basis vectors
are $\{L_{-1}, G^+_{-1/2} G^-_{-1/2},H_{-1}\}$ and  
$\{L_{-1} G^-_{-1/2},  G^-_{-3/2}, H_{-1} G^-_{-1/2}\}$, respectively. 
 The norms obtained from (\ref{Whitnorm}) are
\begin{eqnarray}\label{exN21}
&&\,\,\,{}^{\,}\langle 1| 1\rangle^{\,}\,\,\,=\, \frac{2 c - 2 \Delta + 2 c \Delta - \omega - c \, \omega} {(2 \Delta-\omega) (c-2 \Delta+2 c \Delta-\omega^2)}\,,\\
\label{exN22}
&&{}^-\langle {\ts \frac{3}{2} }| {\ts \frac{3}{2} } \rangle^- = \,   \frac{2^2\,  (2 c^2 - 2 \De + 2 c \, \De -\om + 
 3  c \, \om)}{(2  \De {+}\om) (-2 {+} 2 c {+}  2  \De {+} 3\om) (c {-} 2  \De {+} 2 c  \De {-} \om^2)}\,.
\end{eqnarray}

%%%%%%%%%%%%%%%%%%%%%%%%%%%%%%%%%%%%%
\section{The affine $\widehat{\sll}(2)$ algebra} \label{ssl2}
%%%%%%%%%%%%%%%%%%%%%%%%%%%%%%%%%%%%%

In this  section we briefly review the $\widehat{\sll}(2)$ algebra and its representations, with particular emphasis on the so called relaxed modules. We also define a Whittaker state in the relaxed module and compute its norm for the lowest levels.

%%%%%%%%%%%%%%%%%%%%%%%%%%%%%%%%%%%%%
\subsection{The $\widehat{\sll}(2)$ algebra and its modules} 
%%%%%%%%%%%%%%%%%%%%%%%%%%%%%%%%%%%%%

The affine $\widehat{\sll}(2)$ algebra is spanned by the modes, $J^A_n$, of three dimension-$1$ currents $J^A(z)$ with $A\in\{0,+,-\}$. The  commutation relations that define the untwisted  $\widehat{\sll}(2)$ algebra (with a central extension) are given  by
\begin{eqnarray}\label{sl2algebra}
[J_n^0,J_m^0] = \frac{k}{2} \,n\, \delta_{n+m,0} \,,\quad 
[J_n^0,J_m^{\pm}] = \pm J_{n+m}^{\pm}  \,, \quad
[J_n^+,J_m^-] = 2J_{n+m}^0 + k\,  n\, \delta_{n+m,0} \,,
\end{eqnarray}
where $n, m\in\ZZ$ and $k$ is the level (central charge) of  $\widehat{\sll}(2)$. 

There exist different types of representations of the $\widehat{\sll}(2)$ algebra. 
The standard highest-weight representation is defined by imposing  the following
requirements on the highest-weight state $|j\rb$:
\be\label{standHWV}
J_0^0 |j\rb = j  |j\rb\,,\qquad J_{n>0}^0 |j\rb =J_{n>0}^- |j\rb =J_{n\ge0}^+ |j\rb =0\,,
\ee
and is freely generated by the action of the remaining modes $J_{n}^A$.

A perhaps less known example is the so-called relaxed representation that 
is  obtained by modifying (relaxing) the conditions \eqref{standHWV}. 
The relaxed Verma module will play a central role in the forthcoming discussion, since it is closely related \cite{Feigin:1997} to the (massive) highest weight representation of the $\cN\!=\!2$ superconformal  algebra. 
The primary field $\Phi_j^\la(z)$ in the relaxed module is parametrized by two complex numbers $j$ and $\lambda$.
The corresponding analogue of the highest weight state is defined by the requirements
\begin{eqnarray}\label{relax1}
J_0^0|j,\la\rb\!\!&=&\!\!\lambda |j,\la\rb \,, \non \\
{\bf{J}}^2 | j,\la\rb\!\!&=&\!\!j(j+1) |j,\la\rb\,,
\end{eqnarray}
and by the following relaxed annihilation conditions:
\be \label{relax2}
J^-_{n>0}| j,\la\rb =0 \,,\qquad J^0_{n>0}| j,\la\rb =0 \,,\qquad J^+_{n>0}| j,\la\rb =0\,. 
\ee
The operator ${\bf{J}}^2$ in (\ref{relax1}) is the ordinary $\sll(2)$ quadratic Casimir
\begin{eqnarray}
{\bf{J}}^2={}[(J_0^0)^2 + \half(J_0^+J_0^- + J_0^-J_0^+)] \,.
\end{eqnarray}
We note that equation (\ref{relax1}) implies
\begin{equation} \label{JmJp}
J^-_0 J^+_0 |j,\la\rb =[j(j+1)-\la(\la+1)] |j,\la\rb \,,
\end{equation}
so that when $j=\la$ the relaxed module reduces to the unrelaxed highest weight module. 
The relaxed Verma module $M_{j,\lambda}$ associated with $|j,\la\rb$ is freely generated by the action of the remaining modes of $J^A$
\begin{equation}
M_{j,\lambda}= \mathfrak{U}(\widehat{\sll}_2^-)\otimes \left(\mathfrak{U}(J_0^-)\oplus\mathfrak{U}(J_0^+) \right) |j,\la\rb.
\end{equation}
The level (minus the sum of the mode numbers) defines a  natural grading in the module. Together with the $J_0^0$-charge  it gives the following decomposition
of the relaxed module
 \begin{align}\label{sl2decomp}
M_{j,\lambda}=\bigoplus_{n\in\mathbb{Z}}\bigoplus_{N\ge 0}
 M_{j,\lambda}^{N,n},
\end{align}
where
\begin{align}  
M_{j,\lambda}^{N,n}=\{ |v\rb \in M_{j,\lambda} \mid L_0^{\text{Sug}} |v\rb =(\Delta^{\text{Sug}}+N)| v\rb \,,\,\, 
J_0^0 |v\rb =(\lambda+n)|v\rb\}\,,
\end{align}
and $L^{\text{Sug}}_0$ is the zero mode of the standard Sugawara stress-energy tensor.
The hermitian conjugates of the modes are $(J^0_n)^\dag = J^0_{-n}$ and  $(J_n^-)^\dag = J^+_{-n}$. 
Similar to the $\cN=2$ case the matrix of inner products of descendants (the Shapovalov matrix, $\mathcal{S}$) has a 
block-diagonal structure with respect to \eqref{sl2decomp}.

%%%%%%%%%%%%%%%%%%%%%%%%%%%%%%%%%%%%%%
\subsection{A Whittaker state in the relaxed $\widehat{\sll}(2)$ module}
%%%%%%%%%%%%%%%%%%%%%%%%%%%%%%%%%%%%%%

As for the $\cN\!=\!2$ superconformal algebra, one can define a Whittaker state in the relaxed  $\widehat{\sll}(2)$ module. We define the relaxed  $\widehat{\sll}(2)$  Whittaker state   by the 
following equations\footnote{Just as for the $\cN\!=\!2$ case, an essentially equivalent definition is obtained by interchanging $+\leftrightarrow -$.}
\be \label{sl2Whit}
J_1^- |W\rb = z\, |W\rb^-  , \quad J_{0}^+ |W\rb^- =  2 \,|W\rb \,,
\ee
where all other modes that annihilate $| j,\la\rb$ also annihilate  $|W\rb$ and $|W\rb^-$. Note that the conditions (\ref{sl2Whit}) imply
\be \label{L1Sug}
L^{\text{Sug}}_{1}|W\rb = \frac{2z}{k+2} |W\rb\,.
\ee
The Whittaker state can be decomposed as
\begin{equation} \label{sl2Whitexp}
|W\rb =\sum_{N=0}^{\infty}  z^N  |N,0\rangle\,, \quad |W\rb^- =  \sum_{N=0}^{\infty}  z^{N+1}  |N,-\rangle \,,
\end{equation}
and we call the states $|N,0\rangle$, $|N,-\rangle$ Gaiotto states. 

Just like for the Virasoro case \cite{Gaiotto:2009b,Marshakov:2009}, it is easy to see that the norm of the Whittaker state at a given level is equal to a particular diagonal element of the inverse of the Shapovalov matrix (here $\langle N, -|$ denotes the hermitean conjugate of  $N,-\rb$): 
\begin{eqnarray} \label{sl2Whitnorm}
&&\langle N,0| N,0\rangle \,= 2^{2N}\, \mathcal{S}^{-1} ((J_{-1}^+)^N(J_0^-)^N; (J_{-1}^+)^N(J_0^-)^N)\,,   \nonumber\\
&&\!\! \langle N,-| N,-\rangle = 2^{2N+2}\, \mathcal{S}^{-1}((J_{-1}^+)^N(J_0^-)^{N+1} ;(J_{-1}^+)^N(J_0^-)^{N+1}) \,.
\end{eqnarray}
The Whittaker states satisfy the recursion relations 
\begin{eqnarray}\label{chainsl2}
&& J_{1}^{-}|N,0 \rangle=\,|N{-}1,-\rangle \non \,, \\
&&J^{+}_0 |N,-\rangle\,=\, 2 |N, 0 \rangle \,. 
\end{eqnarray}
Note that when combined with (\ref{sl2Whitexp}) the recursion relations (\ref{chainsl2}) imply (\ref{sl2Whit}). 

We close this section with some examples. At level 1 
the  basis vectors in the sectors with $J_0^0$ charge 0 and -1 are
 $\{J_{-1}^+ J_0^-, J_{-1}^0 , J_{-1}^- J_0^+\}$ and
$\{J_{-1}^+ J_0^- J_0^-$, $J_{-1}^0 J_0^-,$ $J_{-1}^-\}$, respectively. 
The norms are obtained from (\ref{sl2Whitnorm}):
\begin{eqnarray}  \label{exsl21}
&&\,\,\lb 1,0|1,0\rb \,=  \frac{ 2^{2} \, (2 j + 2 j^2 - 2 k - k^2 - 2 \lambda - 2 k \lambda - 
 2 \lambda^2)}{(2 j {-} k) (2 {+} k) (2 {+} 2 j {+} k) (1 {+} 
   j {-} \lambda) (j {+} \lambda)} \,,\\
\label{exsl22}
&&\lb 1,-|1,-\rb =
 \frac{ 2^{4} \, (-2 j - 2 j^2 + k^2 - 2 \lambda + 2 k \lambda + 
 2 \lambda^2) }{(2 {+} k) (2 j {-} k) (2 {+} 2 j {+} k) (2 {+} 
   j {-} \lambda) (1 {+} j {-} \lambda) (1 - 
   j {-} \lambda) (j {+} \lambda)}.\nonumber
\end{eqnarray}

%%%%%%%%%%%%%%%%%%%%%%%%%%%%%%%%%%
\section{$\cN=2$ Gaiotto states as a limit of chain vectors} \label{sGai}
%%%%%%%%%%%%%%%%%%%%%%%%%%%%%%%%%%

Let us recall how the Virasoro Whittaker state arises \cite{Gaiotto:2009b} from a limit of the Liouville four-point conformal block on the sphere, that we schematically write as
\be
\lb V_1| V_2(1) V_3(z) |V_4\rb \,.
\ee
The $V_3(z)V_4(0)$  OPE implies the decomposition 
\begin{equation}
{}[V_3(z)\,|V_4\rb ]_{{\Delta}}\,=\,
z^{\Delta-\Delta_3-\Delta_4}\,\sum_{N=0}^{\infty}z^N |N\rangle_{34}\,,
\end{equation}
where we refer to the $|N\rangle_{34}$ states in the expansion  as chain vectors. The chain vectors satisfy certain recursion relations that specify them completely. 
After the redefinition $|N\rangle_{34}\rightarrow (-\Delta_4)^N |N\rangle_{34}$, 
the Gaiotto states $|N\rb_G$ are defined as the limit of $|N\rb_{34}$, as $\Delta_4$ approaches  $\infty$. Defining the Whittaker state via
\be
|W\rb = \sum_{N=0}^{\infty}z^N |N\rangle_{G}\,,
\ee
the four-point block reduces to $\lb W|W\rb$. 

In the ordinary CFT \cite{Belavin:1984} which has the Virasoro algebra as its chiral symmetry algebra, the conformal symmetry  completely determines  the contribution of the descendants in the OPE. More precisely, conformal symmetry leads to recursion relations for the OPE coefficients, which uniquely specify  some linear combinations of the descendants at the $N$th level in the Verma module contributing in the OPE between any given pair of primary fields. 
In the $\cN\!=\!1$ supersymmetric case the situation becomes more involved (see e.g.~\cite{Belavin:2007} and references therein). Instead of scalar primary fields, there appears super-doublets of primary fields. At the level of the OPE this extension requires one to  independently consider the descendants of the two components of the primary super-doublet.
  
In the $\cN\!=\!2$ supersymmetric case the super-multiplet of primary fields consists of four fields
\begin{align}\label{supmultiplet}
&V_\Delta^\omega\,,\nonumber\\
&V_{\Delta,\omega}^{+}= G_{-1/2}^{+} V_{\Delta}^{\omega}\,,\qquad
V_{\Delta,\omega}^{-}= G_{-1/2}^{-} V_{\Delta}^\omega\,,\\
&\tilde{V}_{\Delta,\omega}=\frac 12 \bigg(G_{-1/2}^{+}G_{-1/2}^{-}-G_{-1/2}^{-}G_{-1/2}^{+}\bigg)V_{\Delta}^\omega\,, \nonumber
\end{align}
and conformal invariance imply four independent channels in the OPE.

The OPE $V_{\Delta_3}^{\omega_3}(z)\,V_{\Delta_4}^{\omega_4}(0)$ involves descendants of the intermediate state $V_{\Delta}^{\omega}$, 
where we denote  the $N$th-level contribution by $|N\rangle_{34}$. The conservation of $H_0$ charge gives the following restriction
\begin{equation}
\omega=\omega_3+\omega_4 + n\,,
\end{equation}
where $n$ is some integer. When this relation is satisfied, the contribution to the OPE comes only 
from the sub-module with $H_0$ charge $\omega+n$: 
\begin{equation}
[V_3(z)|V_4\rb]^{(n)}_{{\Delta}}\,=\,
z^{\Delta-\Delta_3-\Delta_4}\,\sum_{N=0}^{\infty}z^N |N\rangle_{34}\,,
\end{equation}
where $|N\rangle_{34}\in M_{\Delta,\omega}^{N,n}$. We recall that $|N\rangle_{34}=0$ if $N\!<\!n^2/2$.
The states $|N\rangle_{34}$ and the analogous states $|N\rangle^+_{34}$, $|N\rangle^-_{34}$, and $\widetilde{|N\rangle}_{34}$ appearing in the OPEs involving the other components
of the primary super-multiplet \eqref{supmultiplet}
\begin{eqnarray} \label{VVOPEs}
&&[V_3^+(z)|V_4\rb]^{(n+1)}_{\Delta}\,=\,z^{\Delta-\Delta_3-\Delta_4-\frac{1}{2}}
\sum_{N=0}^{\infty} z^N |N\rangle_{34}^+\,,    \nonumber\\
&&[V^{-}_3(z)|V_4\rb]^{(n-1)}_{\Delta}\,=\,z^{\Delta-\Delta_3-\Delta_4-\frac{1}{2}}
\sum_{N=0}^{\infty} z^N |N\rangle_{34}^{-}\,,     \\
&&\,\, [\,\tilde{V}_3(z)|V_4\rb]^{(n)}_{\Delta}\,\,\,\,\,\,=\,z^{\Delta-\Delta_3-\Delta_4-1}
\sum_{N=0}^{\infty} z^N {\widetilde{|N\rangle}}_{34}\,,    \nonumber
\end{eqnarray}
satisfy the following set of recursion relations  \cite{Dorrzapf:1994} 
\begin{eqnarray}\label{chain}
&&H_m\,|N{+}m\rangle_{34}\,=\,\frac{}{}\omega_3 |N\rangle_{34}\,,    \nonumber\\
&&H_m\,{\widetilde{|N{+}m\rangle}}_{34}\,=\,\frac{}{} 2 m\, \Delta_3 |N\rangle_{34}+\omega_3 
{\widetilde{|N\rangle}}_{34} \,,      \nonumber\\
&&H_m\,|N{+}m\rangle_{34}^{\pm}\,=\,\frac{}{}(\omega_3\pm 1)|N\rangle_{34}^{\pm}\,,\nonumber\\
&&G_m^{\pm}|N{+}m\rangle_{34}\,=\,\,\frac{}{}|N\rangle_{34}^{\pm}\,,\\
&&G_m^{\pm}{\widetilde{|N{+}m\rangle}}_{34}\,=
\mp[\Delta+2m(\Delta_3+\frac 12)-\Delta_4\pm(m+\frac 12)\omega_3+N]|N\rangle_{34}^{\pm}\,,\nonumber\\
&&G_m^{\pm}|N{+}m\rangle_{34}^{\mp}\,=\,\,[\Delta+2 m \Delta_3-\Delta_4 \pm (m+\frac 12)\omega_3 +N]|N\rangle_{34}\pm
\widetilde{|N\rangle}_{34}\,,\nonumber\\
&&G_m^{\pm}|N{+}m\rangle_{34}^{\pm}\,=\,\frac{}{}0\,,\nonumber
\end{eqnarray}
where $m\!>\!0$. For convenience we also write the action of the generators $L_m$, even though the result follows directly from \eqref{chain}
\begin{eqnarray}
&&\,L_m\,|N{+}m\rangle_{34}\,=\,\frac{}{}[\Delta+ m\Delta_3-\Delta_4+N]|N\rangle_{34}\,,\nonumber\\
&&\,L_m\,{\widetilde{|N{+}m\rangle}}_{34}\,=\frac{m(m+1)\omega_3}{2}{|N\rangle}_{34}+[\Delta+m(\Delta_3+1)-\Delta_4+N] {\widetilde{|N\rangle}}_{34}\,,\;\;\;\\
&&\,L_m\,|N{+}m\rangle_{34}^{\pm}\,=\,\frac{}{}[\Delta+m(\Delta_3+\frac 12)-\Delta_4+N] |N\rangle_{34}^{\pm}\,.\nonumber
\end{eqnarray}
 The set of recursion relations (\ref{chain}) do not determine the chain vectors completely. In particular, the ratio of $|N\rangle_{34}$ and ${\widetilde{|N\rangle}}_{34}$ is not determined. 
 
  By rescaling,  
 \be
 |N\rb_{34} \rar (-\Delta_4)^N |N\rb_{34} \,, \quad |N\rb^\pm_{34} \rar 
(-\De_4)^{N+\frac{1}{2}} |N\rb^\pm_{34} \,, \quad \widetilde{|N\rangle}_{34} \rar (-\Delta_4)^{N+1} \widetilde{|N\rangle}_{34}
\ee
we find that in the Gaiotto limit where $\De_4$ approaches $\infty$, the recursion relations (\ref{chain}) reduce to 
\begin{eqnarray}\label{GaiottochainN2}
&&G_{\frac 12}^{\pm}|N{+}\half\rangle\,=\,\,|N \rangle^\pm\, , \nonumber\\
&&G_{\frac 12}^{\pm}|N{+}\half \rangle^\mp= |N\rangle \pm \widetilde{|N\rangle}\,,\nonumber\\
&&G_{\frac 12}^{\pm}\widetilde{|N{+}\half\rangle}\,=\,\, \mp|N \rangle^\pm \,.
\end{eqnarray}
It turns out that this system can be diagonalised (which is apparently 
not the case in the general massive case). 
To see this, we introduce the ``symmetric'' and ``antisymmetric'' combinations
\begin{eqnarray}
|N\rangle_s=|N\rangle+\widetilde{|N\rangle}\,, \nonumber\\
|N\rangle_a=|N\rangle-\widetilde{|N\rangle}\,.
\end{eqnarray} 
When these relations are inserted into \eqref{GaiottochainN2} we find that the recursion relations can be separated in two completely decoupled  parts
\be
G_{\frac 12}^{+}|N{+}\half\rangle_a\,=-2|N\rangle^{+} \,, \quad 
G_{\frac 12}^{-}|N{+}\half\rangle^+\,= |N\rangle_{a}\,,    \quad
G_{\frac 12}^{-}|N{+}\half\rangle_a\,=0\,,
\ee
and
\be \label{BPS}
G_{\frac 12}^{-}|N{+}\half\rangle_s\,=\,\,\frac{}{}2|N\rangle^{-} \,, \quad
G_{\frac 12}^{+}|N{+}\half\rangle^-\,= |N\rangle_{s}. \quad
G_{\frac 12}^{+}|N{+}\half\rangle_s\,= 0 \,.
\ee
Since the two sectors are completely decoupled and essentially equivalent we can without loss of generality restrict ourselves to one of the two sectors. In the remainder of the paper we focus on (\ref{BPS}). For simplicity we drop the subscript $s$ on $|N\rb$ in other sections.

%%%%%%%%%%%%%%%%%%%%%%%%%%%%%%%%%%%%%%
\section{Relating blocks for $\cN\!=\!2$, relaxed $\widehat{\sll}(2)$ and ordinary $\widehat{\sll}(2)$ }\label{sRel}
%%%%%%%%%%%%%%%%%%%%%%%%%%%%%%%%%%%%%%

The goal of this section is to derive relations between the irregular conformal blocks in $\cN\!=\!2$, relaxed $\widehat{\sll}(2)$ and ordinary $\widehat{\sll}(2)$. We first discuss the relation between the $\cN\!=\!2$ and relaxed $\widehat{\sll}(2)$ modules. Then in \ref{relblocks} we generalise this relation to the conformal blocks and in particular show the equivalence between the $\cN\!=\!2$ and  relaxed  $\widehat{\sll}(2)$  irregular blocks. Finally in \ref{sanal} we show that the irregular relaxed  $\widehat{\sll}(2)$ blocks can be obtained from the irregular ordinary $\widehat{\sll}(2)$ blocks by means of a certain well-defined analytic continuation procedure that we explicitly describe.

%%%%%%%%%%%%%%%%%%%%%%%%%%%%%%%%%%%%%&
\subsection{Relating $\cN\!=\!2$ and relaxed $\widehat{\sll}(2)$ modules via coset construction \label{relmods}}
%%%%%%%%%%%%%%%%%%%%%%%%%%%%%%%%%%%%%%

The $\cN\!=\!2$ superconformal algebra can be realised in terms of $\widehat{\sll}(2)$ and a complex fermion $\psi(z)$ ($\bar\psi(z)$) by using the coset construction 
of Di Vecchia {\it et al.}~\cite{DiVecchia:1986}.   This construction is a particular case of the so-called Kazama--Suzuki construction \cite{Kazama:1988}. More precisely, the $\cN\!=\!2$ superconformal algebra is the symmetry algebra of the coset
\be \label{21coset}
\frac{\hat{\sll}(2)_k\! \times\widehat{\ul}(1)}{\widehat{\ul}(1)} \,,
\ee 
where $k$ is the level of the affine $\widehat{\sll}(2)$ algebra. 
To describe the details of this relation it is convenient to use the OPE language. The complex fermion and the $\widehat{\sll}(2)$ current algebra satisfy the OPEs
\bea
&&\; \; \; \psi(z) \bar\psi(w) \sim \frac{1 }{(z-w)}\,,  \qquad \quad 
J^0(z)J^0(w)\sim \frac{{\ts \frac{k}{2} }\, \delta_{ab}}{(z-w)^2} \,, \non \\ 
&&J^0(z)J^{\pm}(w)\sim \pm \frac{J^{\pm} (w)}{(z-w)}\,, \qquad  
J^+(z)J^-(w)\sim \frac{k }{(z-w)^2}+ \frac{2 J^0(w)}{(z-w)}\,,
\eea
and the non-singular OPEs of the $\cN\!=\!2$ algebra are 
\begin{eqnarray}\label{N2ope}
&& T(z)T(w)=\frac{3c/2}{(z-w)^4}+\frac{2}{(z-w)^2}T(w)+\frac{1}{(z-w)}\pa T(w)  \,,\nonumber\\
&& T(z)H(w)=\frac{1}{(z-w)^2}H(w)+\frac{1}{z-w}\pa H(w)  \,, \nonumber\\
&& T(z)G^{\pm}(w)=\frac{3/2}{(z-w)^2}G^{\pm}(w)+\frac{1}{z-w}\pa G^{\pm}(w)\,, \\
&& H(z)H(w)=\frac{c}{(z-w)^2} \,, \qquad H(z)G^{\pm}(w)=\pm\frac{1}{(z-w)}G^{\pm}(w) \,,
\nonumber\\
&& G^{\pm}(z)G^{\mp}(w)=\frac{2c}{(z-w)^3}\pm\frac{2}{(z-w)^2}H(w)+\frac{1}{z-w}(2T(w)\pm \pa H(w)) \,. \nonumber
\end{eqnarray}

The $\widehat{\ul}(1)$ algebra in the numerator of the coset (\ref{21coset}) corresponds to the
conserved current $\psi \bar \psi(z)$ 
while the $\widehat{\ul}(1)$ in the denominator is the diagonal sub-algebra generated by 
\be  \label{Kdef}
K(z)=J^0(z)-\psi\bar\psi(z)\,.
\ee
It is straightforward to verify that the odd generators $G^{\pm}$ constructed as
\begin{equation}  \label{QGsl}
  G^+(z)= \sqrt{\frac{2}{k{+}2}}\, \psi(z) J^+(z)\,,\qquad 
  G^-(z)= \sqrt{\frac{2}{k{+}2}}\, \bar \psi(z) J^-(z)\,,
\end{equation}
have dimension $3/2$ and have vanishing OPEs with $K(z)$. Furthermore, the dimension 1 current $H(z)$ and the stress-energy tensor $T(z)$ of the $\cN=2$ algebra are  uniquely fixed from the $G^+(z)G^-(w)$ operator product expansion: 
\begin{eqnarray} \label{HTJ}
&&H(z)=\frac{1}{2(k+2)} J^0(z)+\frac{k}{k+2}\psi\bar \psi(z) \,, \\
&&T(z)=\frac{1}{2(k+2)}\big(J^+J^-(z)+J^-J^+(z)\big)+\frac{k}{k+2}\psi \bar\psi(z)\nonumber
+\frac{2}{k+2}J^0 \psi\bar\psi (z)\,.
\end{eqnarray}
The $G^+(z)G^-(w)$ OPE also fixes the normalisation of $G^\pm(z)$.
The central charge of the $\cN\!=\!2$ algebra is expressed in terms of the $\widehat{\sll}(2)$ level as 
\begin{equation}\label{ck}
c={\frac{k}{k+2}}\,.
\end{equation}

To construct the highest weight representation we need to specify the primary fields 
of the $\cN\!=\!2$ algebra. They can be constructed from the relaxed $\widehat{\sll}(2)$ primary fields dressed by the exponents $e^{\bet \phi}$ for a suitably chosen $\bet$ 
\be
V^{\omega}_{\Delta}(z)=\Phi_j^{\lambda}(z) e^{\bet\phi}(z)\,,
\ee
where the bosonic field $\phi$ is related to the current $K(z)$ as
\be
K(z)=\partial \phi(z)\,.
\ee
The parameter $\bet$ is fixed by the requirement of having vanishing OPEs with $K(z)$. This requirement also fixes the relations between the parameters of the primary fields (which also follow from (\ref{HTJ}))
\be \label{dj}
\Delta=\frac{j(j{+}1)-\lambda^2}{k+2}\,,\qquad  \omega=\frac{2\lambda}{k+2}\,.
\ee

The basic tool to compare the structure of the representations 
  is the evaluation of the characters.   
The character of the  irreducible relaxed $\widehat{\sll}(2)$ representation $M_{j,\lambda}$, defined modulo the contribution of the zero modes $J_0^{\pm}$,  is given by
\begin{equation}
  \chi_{j,\lambda}^{>0}(z,q)=
  {\rm Tr}_{M_{j,\lambda}}(q^{L^{\rm Sug}_0}\,z^{J^0_0})=
  \frac{z^\lambda\,q^{\Delta_j}}
 {\prod_{i=1}^\infty(1-zq^i)\,
    \prod_{i=1}^\infty(1-z^{-1}q^i)\,
   \prod_{i=1}^\infty(1-q^i)}\,,
\end{equation}
where $\Delta_j$ is the Sugawara dimension and the grading related to the $z$ variable defines 
a sector with a given value of the $J_0^0$ charge. The additional action of the $J_0^{\pm}$ generators in the relaxed module allows one to change this charge without changing the level. Effectively this means that one should sum over all sectors. Hence, we find the character of the relaxed module at some fixed charge $J_0^0=\lambda+n$
\begin{equation}\label{SL2char}
  \chi_{j,\lambda}^{\rm sl2}(z,q)=\chi_{j,\lambda}^{ > 0}(1,q)
  \sim \chi_{\rm boson}(q)^3=1+3q+9q^2+22q^3+51q^4+\dots
\end{equation}
We note that the number of states in the relaxed module does not depend on $n$.
In \eqref{SL2char} $\chi_{\rm boson}$ is the free boson character 
\be \label{chibos}
 \chi_{\rm boson}(q) =\frac{1}{\prod_{i=1}^\infty(1-q^i)}  \,.
\ee

The irreducible character of the highest weight $\cN\!=\!2$ module is
\begin{equation}  \label{chiN2}
  \chi_{\Delta,\omega}^{\cN=2}(z,q)=
  {\rm Tr}_{M_{\Delta,\omega}}(q^{L_0}\,z^{H_0})=
  \frac{z^{\omega} \,q^{\Delta}\, \prod_{k>0}(1+zq^{k-\frac 12})\prod_{k>0}^\infty(1+z^{-1}q^{k-\frac 12})}
 {\prod_{k>0}(1-q^k)\prod_{k>0}(1-q^k)}\,.
\end{equation}
Applying the Jacobi triple-product identity this expression can be written in the following form
\begin{equation}
  \chi_{\Delta,\omega}^{\cN=2}(z,q)=
  \frac{z^{\omega} \,q^{\Delta}\, \sum_{n\in \mathbb{Z}} z^n q^{\frac{n^2}{2}}}
 {\prod_{k>0}(1-q^k)^3}\sim
 \sum_{n\in \mathbb{Z}} z^n q^{\frac{n^2}{2}}\chi_{\rm boson}(q)^3\,,
\end{equation}
Thus, up to a shift of the lowest level, the number of states in the $\cN\!=\!2$ highest weight module with a fixed value of the $H_0$ charge coincides with the number of states in the relaxed  $\widehat{\sll}(2)$ module with an arbitrary fixed value of the $J_0^0$ charge. 

%%%%%%%%%%%%%%%%%%%%%%%%%%%%%%%%%%%%
\subsection{Equivalence between $\cN\!=\!2$ and relaxed $\widehat{\sll}(2)$ irregular blocks }  \label{relblocks}
%%%%%%%%%%%%%%%%%%%%%%%%%%%%%%%%%%%%

In this subsection we show that the irregular $\cN\!=\!2$ and relaxed $\widehat{\sll}(2)$ conformal blocks are equal, i.e.~the norms of the Whittaker states \eqref{Whitexp} and \eqref{sl2Whitexp} are equal
\begin{equation}
 \langle W|W\rangle_{\cN=2}=  \langle W|W\rangle_{\widehat{\sll}(2)}\,, \qquad  {}^-\langle W|W\rangle^-_{\cN=2}=  {}^-\langle W|W\rangle^-_{\widehat{\sll}(2)}\,,
\end{equation}
provided we also redefine the variable $z$ used in the 
two theories as 
\be \label{zztrans} 
z_{\hat{\sll}(2)}=\frac{k+2}{2}\,z_{\cN=2} \,.
\ee 
The reason for this rescaling is a result of our normalisations and can be traced to (\ref{L1Sug}). In terms of the Gaiotto states the above statement reads (here $N$ is an integer)
\begin{equation}\label{equiv}
 \langle N|N\rangle_{\cN=2}=  \langle N,0|N,0\rangle_{\widehat{\sll}(2)}\,, \qquad  {}^-\langle N{+}\half |N{+}\half\rangle^-_{\cN=2}=  \langle N,- |N,-\rangle_{\widehat{\sll}(2)}\,.
\end{equation}
It is straightforward to check these relations for $N\!=\!1$ by comparing (\ref{exN21}), (\ref{exN22})  with (\ref{exsl21}), (\ref {exsl22}) using (\ref{dj}). However, we will in fact show the equalities (\ref{equiv}) for any $N$ by showing that the Gaiotto states in the two theories are mapped into each other under the coset map discussed in section \ref{relmods}.

As discussed in section \ref{sGai}, the chain vector in the $\cN\!=\!2$ Verma module is a state which appears in the OPE $V_3(z)V_4(0)$ of two $\cN\!=\!2$ primary fields.
Using the coset map discussed in section \ref{relmods} it can be written as
\begin{equation}\label{chainFac}
|N\rangle_{34}=\sum_{m=0}^N |N{-}m,0\rangle_{34} |m\rangle_{\Psi}\,.
\end{equation} 
Here, in each term in the sum,  the first factor is a chain vector associated with the OPE of two  relaxed  $\widehat{\sll}(2)$  primary fields $\Phi_3(z) \Phi_4(0)$, whereas the second factor is a free fermion chain vector which appears in the OPE of the bosonic dressing  exponents $e^{\bet_3 \phi(z)}e^{\bet_4 \phi(0)}$. Since the two sectors do not interact we have the form \eqref{chainFac}. As discussed in section \ref{sGai} the Gaiotto states arise as a limit of the chain vectors. 

The $\cN\!=\!2$ Gaiotto states $|N\rangle$, $|N{+}\half \rangle^-$ are uniquely fixed by the recursion relations we wrote down in section \ref{sN2} 
\begin{eqnarray}\label{4chainN2}
&&G_{\frac 12}^{-}|N\rangle\,=\,\,|N{-}\half \rangle^-, \, \nonumber\\
&&G_{\frac 12}^{+}|N{+}\half \rangle^-\,=\,\,2 |N\rangle \,.
\end{eqnarray}
Here the action of all other positive modes annihilate the states. Note from the algebra \eqref{N2alg} that $G_{1/2}^\pm$ and $H_1$ generate all positive modes. 

The relaxed  $\widehat{\sll}(2)$ Gaiotto states $|N,0 \rangle$, $|N{-}1,-\rangle$ are uniquely fixed by the recursion relations we wrote down in section \ref{ssl2}
\begin{eqnarray}\label{4chainsl2}
&& J_{1}^{-}|N,0 \rangle=\,|N{-}1,-\rangle \non \,, \\
&&J^{+}_0 |N,-\rangle\,=\, 2 |N, 0 \rangle \,. 
\end{eqnarray}
Here the action of all other non-negative modes annihilate the states. Note from the algebra \eqref{sl2algebra} that $J_{0}^\pm$ and $J^0_1$ generate all non-negative modes.

Hence, the equivalence (\ref{equiv}) is proven if  the recursion relations for the $\cN\!=\!2$ Gaiotto states follow from the relations for the $\widehat{\sll}(2)$ Gaiotto states and the relations for the free fermion chain vectors.

In the free fermion sector, the Gaiotto limit of the chain vector is the vacuum state $|0\rangle_\Psi$, since $\Psi_r^\pm |N\rangle_\Psi=0$ for $r=\frac{1}{2}, \frac{3}{2},\ldots$. Thus, only the first term in the sum \eqref{chainFac} contributes and we find that the $\cN\!=\!2$ Gaiotto state has the factorised form
\begin{equation}\label{N0}
|N\rangle=|N,0\rangle|0\rangle_\Psi\,.
\end{equation}
With this result we are in a position to check whether the recursion relations
\eqref{4chainsl2} lead to \eqref{4chainN2}. 
We only need the first few terms in the explicit form of the map \eqref{QGsl}
\begin{eqnarray}\label{generators}
&&G_{1/2}^+=J_1^+\Psi_{-1/2}^+ + J_0^+\Psi_{1/2}^+ +...\nonumber\\
&&G_{1/2}^-=J_1^-\Psi_{-1/2}^- + J_0^-\Psi_{1/2}^- +...\nonumber\\
&&H_1=J_1^0+\Psi_{1/2}^+\Psi_{1/2}^-+...
\end{eqnarray}
and also for the the current $K(z)$ defined in \eqref{Kdef}
\begin{equation}\label{Kcurrent}
K_1=J_1^0-\Psi_{1/2}^+\Psi_{1/2}^-+...
\end{equation}
We note that by construction the $\cN\!=\!2$ sector commutes with $K(z)$, so all the  
$\cN\!=\!2$ Gaiotto states should be annihilated by $K_1$. 
It is easily verified using the properties of the $\widehat{\sll}(2)$ Gaiotto state $|N,0\rangle$ that the $\cN\!=\!2$ Gaiotto state \eqref{N0} is annihilated  by $K_1$ as well as by $H_1$ and $G_{1/2}^+$. 
 
Using the second equation in \eqref{generators}, we find from the first equations in  \eqref{4chainN2} and \eqref{4chainsl2} that
\begin{eqnarray}
|N{+}\half\rangle^-=|N,-\rangle \Psi_{-1/2}^- |0\rangle_\Psi\,.
\end{eqnarray} 
The state $|N\rangle^-$ is annihilated by $K(z)$ since it is obtained
by the action of $G^-_{1/2}$ on $|N\rangle$ and $K(z)$ commutes with the $\cN\!=\!2$ algebra.
It is also annihilated by $H_1$ and $G_{1/2}^-$. This follows the properties of the $\widehat{\sll}(2)$ Gaiotto state $|N,-\rangle$, together with the obvious requirement $\Psi_{-1/2}^2=0$. 

Finally, by acting with $G_{1/2}^+$ and using the second equation in \eqref{4chainsl2} we find
\be
G_{1/2}^+|N{+}\half \rb=J_0^+|N,0\rangle \Psi_{+1/2}^+\Psi_{-1/2}^-|0\rangle_\Psi 
=2 |N,0\rangle |0\rangle_\Psi = 2|N \rb\,,
\ee
which is the second equation in \eqref{4chainN2}. Thus, we have shown that $|N\rb$, $|N{+}\frac{1}{2}\rb$ satisfy the $\cN\!=\!2$ recursion relations \eqref{4chainN2} provided $|N,0\rangle$, $|N,-\rb$ satisfy the $\widehat{\sll}(2)$ recursion relations \eqref{4chainsl2}, which proves \eqref{equiv}.

%%%%%%%%%%%%%%%%%%%%%%%%%%%%%%%%%%%%%%%%%%%%%%%%%%%%%%%%%%%%%%%%%%%%%%%%%%%%%%%%

%%%%%%%%%%%%%%%%%%%%%%%%%%%%%%%%%%%%%%%%%%%%
\subsection{Relating ordinary and relaxed $\widehat{\sll}(2)$ blocks via analytic continuation} \label{sanal}
%%%%%%%%%%%%%%%%%%%%%%%%%%%%%%%%%%

In this subsection we show that the ordinary  (unrelaxed)  $\widehat{\sll}(2)$ irregular conformal blocks are related to the irregular conformal blocks associated with the relaxed $\widehat{\sll}(2)$ module via a certain analytic continuation.  

This relation follows from the following arguments. Consider the basis of descendants with level $m$ and ($J_0^0$) charge $s$ in the relaxed module. Denote the basis elements by $B_i^{(s)}|j,\la\rb $, where $B_i^{(s)}$ denotes a string of $J^A_n$'s and $i$ labels the different possible choices. The basis elements can be constructed as follows. First construct all possible descendants with level $m$ and any charge out of only non-zero-modes generators (i.e.~$J^A_n$'s with $n>0$). Each of these descendants can be used to construct a descendant at level $m$ with the right charge $s$ by including either an additional number of $J_0^-$'s or a number of $J_0^+$'s. Note that using both $J_0^-$ and $J_0^+$ at the same time does not lead to independent basis elements, since a $J_0^-$ $J_0^+$ pair can be removed by the eigenvalue properties of the relaxed module (\ref{JmJp}). 

Next consider the unrelaxed case at same level but with charge $s - n$ where $n$ is much larger than $|s|$.  We first construct descendants with the right level out of non-zero modes as above. Since $n\gg |s|$ each of these descendants can be used to construct a descendant with charge  $s - n$ by including a suitable number of additional $J_0^-$'s. However, we can  alternatively also use descendants of the form $B_i^{(s)} (J_0^-)^n|j\rb$ as our basis elements, where $B_i^{(s)}$ is exactly the same string of $J^A_n$'s as in the relaxed module. This is simply a linear change of basis compared to the basis involving only $J_0^-$'s. This follows from the fact that all $J_0^+$'s present in $B_i^{(s)}$ can be moved to the right and annihilated against the highest weight state $|j\rb$, leaving a linear combination of terms involving only non-zero modes and $J_0^-$'s. 

To conclude, we can choose the basis elements of descendants in the two cases to be $B_i^{(s)}|j,\la\rb $ (relaxed) and $B_i^{(s)} (J_0^-)^n|j\rb$  (unrelaxed) with the same $B^{(s)}_i$ in both cases.

For both the relaxed and unrelaxed $\widehat{\sll}(2)$  modules, the matrix of inner products 
 of descendants (the Shapovalov matrix) has a block-diagonal structure, where each block 
contains only descendants with a given level $m$ and with the same value of the charge $s$. 
Using the above choices of basis elements, it is easy to see that after eliminating the non-zero modes using the algebra and the properties of the modules, each entry in the Shapovalov matrix 
reduces to a sum of terms of the form (the coefficients depend on the
level and are the same in both cases since the algebra is the same, and $r$ and $q$ are some integers):
\begin{eqnarray}
 \langle j, \lambda |(J_0^+)^q (J_0^0)^r (J_0^-)^q | j, \lambda \rangle =
(\lambda-q)^r \prod_{\ell=0}^{q-1} (j-\lambda+\ell+1)(j+\lambda-\ell),
\end{eqnarray}
for the relaxed module and
\begin{eqnarray}
\langle j |(J_0^+)^{q+n} (J_0^0)^r (J_0^-)^{q+n} | j \rangle =   M(n)\, 
(j-n-q)^r  \prod_{\ell=0}^{q-1} (n+\ell+1)(2j-n-\ell),
\end{eqnarray}
for the unrelaxed module, where (cf.~\cite{Kozcaz:2010b})
\begin{equation} \label{Mn}
M(n)\equiv \langle j|(J_0^+)^n (J_0^-)^{n}|j\rangle=(-2j)_n n!(-1)^n.
\end{equation}
If one now replaces $n$ by $j-\lambda$ in the unrelaxed case with charge $s{-}n$ and
normalises the result by dividing by $M(n)$ one obtains the relaxed
result with charge $s$ at the same level. 
Since the norm of the Whittaker vector is given in terms of a fixed component of the inverse of the Shapovalov matrix, cf.~(\ref{sl2Whitnorm}), we conclude from the above analysis that the coefficient at order $z^m$ of the norm of the Whittaker vector in the relaxed case can be obtained from the coefficient at order  $x^n z^m$ (with $n$ large) of the norm of the Whittaker vector in the unrelaxed case by multiplying by $M(n)$ and replacing $n$ by $j{-}\lambda$.

%%%%%%%%%%%%%%%%%%%%%%%%%%%%%%%%%%%%%%%%%
\section{Surface defects and instanton partition functions}\label{sSurf}
%%%%%%%%%%%%%%%%%%%%%%%%%%%%%%%%%%%%%%%%%

The dual gauge theory description of $\widehat{\sll}(N)$ conformal blocks involves $\cN\!=\!2$ $\SU(N)$ gauge theories with a certain surface defect \cite{Braverman:2004a,Alday:2010,Kozcaz:2010b}. The relevant  instanton moduli space is known as an affine Laumon space \cite{Feigin:2008}. It can equivalently be viewed as an orbifold of the standard ADHM instanton moduli space, and in this language is referred to as a chain-saw quiver \cite{Finkelberg:2010}.  The instanton partition function was computed using the affine Laumon space language in \cite{Feigin:2008}, and using the orbifold language in \cite{Kanno:2011}. 

The surface defect associated with the $\widehat{\sll}(N)$ algebra belongs to a more general class of surface defects in $\SU(N)$ gauge theories that are classified by  partitions of $N$ \cite{Gukov:2006}. The $\cN\!=\!2$ $\SU(N)$ gauge theories with a surface defect belonging to this class have been conjectured \cite{Braverman:2010,Wyllard:2010a} to be associated with  the chiral symmetry algebras that are obtained by quantum Drinfeld-Sokolov reduction from $\widehat{\sll}(N)$. Such algebras are also classified by partitions of $N$. The instanton partiton functions for the case corresponding to a general partition of $N$ were conjectured in \cite{Wyllard:2010b} and confirmed in \cite{Kanno:2011}. The instanton moduli spaces corresponding to a general partition of $N$  can also be formulated in two ways. The analysis in  \cite{Kanno:2011}  uses the orbifold  chain-saw quiver language that we also use in this paper. 

Using complex coordinates, the chain-saw quiver orbifold  acts in the spacetime of the gauge theory as $\CC {\times}(\CC/\ZZ_p)$, where $(z_1,z_2)\rar (z_1,\om z_2)$ with $\omega=\exp(\frac {2\pi i}{p})$. Here $p$ is the length of the partition of $N$ specifying the surface defect. The surface defect is located at $z_2\!=\!0$. Although our main interest is in the case $N\!=\!p\!=\!2$ we first discuss the general case before  specialising to the case of main interest to us. 

In this paper  we formulate the result for the instanton partition functions in a slightly different (but equivalent) way compared to the approach in \cite{Kanno:2011}. Our approach follows closely the one used in  \cite{Fucito:2004b} for  another $\ZZ_p$ orbifold of $\CC^2$, corresponding to the $A_{p-1}$ ALE space. For that case the orbifold acts as  $(z_1,z_2)\rar (\om z_1,\om^{-1} z_2)$ with $\omega=\exp(\frac {2\pi i}{p})$.

We first recall the main points of the ADHM construction of the instanton moduli space, then describe how the orbifold modifies this space,  
and finally express the instanton partition function for the $\cN\!=\!2$ $\SU(N)$ gauge theory in the presence of a surface defect in terms of coloured Young diagrams.

The general $k$-instanton solution 
in the four-dimensional euclidean $\SU(N)$ theory is given by means of the ADHM construction in terms of the matrices $B_1$, $B_2$, $I$, $J$ with the corresponding dimensions $k{\times} k$, $k{\times} k$, $k{\times} N$ and $N{\times} k$. Introducing two vector spaces $V$ and $W$
with $\dim_{\mathbb C}V=k$ and $\dim_{\mathbb C}W=N$, we have
 $I\in \text{Hom}(V,W)$, $J\in \text{Hom}(W,V)$ and  $B_{1,2}\in \text{End}(V)$.  These matrices 
satisfy  
\begin{equation}\label{ADHM}
[B_1,B_2]+I J=0\,.
\end{equation}
The  instanton moduli space $\mathcal{M}_N(k)$ is the space of $\GL(k,\CC)$ gauge orbits in the space of  solutions to (\ref{ADHM}). The $\GL(k,\CC)$  gauge transformation acts as
\begin{eqnarray}
B_{1,2}\rightarrow G B_{1,2} G^{-1},\quad I\rightarrow G I, \quad  J\rightarrow J G^{-1},
\end{eqnarray} 
where $G\in \GL(k,\CC)$. 

The additional restrictions on the ADHM data imposed by the  $\mathbb Z_p$ orbifold are encoded in the equations
\begin{eqnarray}\label{Z2constr}
&& B_1= \ga B_1 \ga^{-1}\,, \qquad \omega B_2=\ga B_2 \ga^{-1}\,,\\
&&\;\;\, I= \ga \, I \, \Ups\,, \qquad \quad \;\;  \omega J= \Ups^\dagger \, J \, \ga^{-1}\,,\nonumber
\end{eqnarray}
where $\omega=\exp(\frac {2\pi i}{p})$, $\ga\in \GL(k)$ and $\Ups \in \U(N)$. We note that this system is consistent with 
the ADHM constraint \eqref{ADHM}. From \eqref{Z2constr} it follows that $\ga^p=\id_{k}$ and $\Ups^p=\id_{N}$, which implies that $\ga$ and $\Ups$ can be chosen to be diagonal matrices with the structure 
\be
\ga=
\left(\begin{array}{c|ccc|c}
 \omega \,\id
 _{\small{k_1}}  &     &0 & & 0 \\
\hline
  0   &            &   \ddots    &   & 0  \\
\hline
   0 &            &       0        & & \omega^{p}\,\id_{\small{k_{p}}}  
\end{array}\right) \!,   
\qquad
\Ups=
\left(\begin{array}{c|ccc|c}
  \omega\,\id_{\small{N_1}} \,\, &     &0 & & 0 \\
\hline
  0   &            &   \ddots    &   & 0  \\
\hline 
   0  &            &       0        & & \omega^{p}\,\id_{\small{N_{p}}}\,\, 
\end{array}\right) \!,
\ee
where
\begin{equation}
\sum_{L=1}^{p} k_L= k,\qquad \sum_{L=1}^{p} N_L=N.
\end{equation}
The second equation defines a partition\footnote{More precisely it defines a composition of $N$. Different compositions related to the same partition correspond to the same surface defect but lead to different expressions for the instanton partition functions, see~\cite{Kanno:2011} for a discussion.} of $N$ that is the partition that specifies the type of surface defect.  
It will prove useful to introduce the notation $q_\alpha$, $\alpha\in 1,...,N$ for the eigenvalues 
of $\Ups$
so that $\Ups_{\alpha,\alpha'}=  \omega^{q^\alpha} \delta_{\alpha,\alpha'}$. 
Each pair $(\{k_L\},\{q_\alpha\})$ is related to some connected component of the moduli space. 
We denote the corresponding connected component $\mathcal{M}_{N,p}(\{k_L\},\{q_\alpha\})$.
Note that if some $N_L$ is zero the corresponding $\{q_\al\}$ component does not have an immediate surface defect interpretation (we will return to this point later). 
The equations \eqref{Z2constr} imply a block structure for the matrices $B_1$, $B_2$. For example, from $[B_1,\ga]=0$, it follows
that $B_1$ is block diagonal with blocks of the same sizes as in the matrix $\ga$.   
If we label the blocks by pairs $(L_1,L_2)$ we get
\begin{equation}
B_1^{(L_1L_2)}\sim \delta_{L_1,L_2}, \qquad B_2^{(L_1L_2)}\sim \delta_{L_1+1,L_2\text{ mod }p}.
\end{equation} 
When $p\!=\!2$, $B_{1,2}$ are $2{\times}2$ block matrices, with $B_1$ block diagonal and $B_2$ block off-diagonal. 

Next we describe the fixed points of the vector field 
induced by the torus action $T$ on the manifold  $\mathcal{M}_{N,p}(\{k_L\},\{q_\alpha\})$
\begin{equation}\label{torus-action}
T:\quad B_1\mapsto t_1 B_1 ; \, \, \, \, B_2\mapsto t_2 B_2 ; \, \, \, \, I_{i\al} \mapsto I_{i\alpha}t_\alpha; \, \, \,
\, J_{\al i}\mapsto t_1 t_2 t_{\alpha}^{-1}J_{\alpha i},
\end{equation}
where $t_{1}=e^{\epsilon_{1} }$, $t_{2}=e^{\epsilon_{2} }$, $t_\alpha=e^{a_{\alpha}}$ and $\prod_{\alpha=1}^r t_\alpha=1$.  A statement that is important for our considerations is that equation \eqref{Z2constr}, defining the orbifold, is in fact the requirement of symmetry with respect to a 
particular torus element, namely: $t_1\!=\!1$, $t_2\!=\!\omega$ and $t_\al\!=\!\Ups_{\al\,\al}$. This means that we can alternatively think of the orbifolding as acting in the following way 
\be \label{orbaction}
a_\al \rar a_\al + q_\al \frac{2\pi i}{p}  \qquad \ep_2 \rar \ep_2 + \frac{2\pi i}{p} \qquad \ep_1 \rar \ep_1\,.
\ee
All fixed points of the torus action automatically belong to the $\mathbb{Z}_p$ symmetric subspace \eqref{Z2constr}.  The fixed points of the torus action \eqref{torus-action} are well known \cite{Nekrasov:2002,Flume:2002}, and are labelled by $N$-tuples of Young diagrams $\vec{Y}=(Y_1,\dots,Y_{N})$. 

 The  character $\chi$ of the tangent space   
at a fixed point ($\chi=\sum_i \Lambda_i$, where $\Lambda_i$ are the eigenvalues of $T$) before orbifolding  is given by 
\be \label{usualchar}
\chi(t_1,t_2,e_{\alpha})= -V^* \, V (t_1-1)(t_2-1)  + W^*\, V+ V^*\,W \, t_1 t_2 \,,
\ee
where
\be \label{VW}
V= \sum_{\al=1}^{N} \sum_{j_\al=1}^{\ell_\al} \sum_{i_{\al}=1}^{\nu_{j_{\al}} } e^{a_{\al}}  e^{-\ep_1(i_\al-1)} e^{-\ep_2(j_\al-1)} \,, \qquad W=\sum_{\al=1}^N e^{a_{\al}}\,,
\ee
and the superscript $^*$ means change sign of all terms in the exponents. The indices $i_\al$ and $j_\al$ label the rows and columns of $Y_{\al}$, $\nu_{j_\al}$ denotes the number of boxes in column $j_\al$, and $\ell_\al$ denotes the number of columns. 
The character (\ref{usualchar}) can also be rewritten 
\be \label{char}
\sum_{\alpha,\beta=1}^N 
\sum_{s_\al \in Y_\alpha} \left[ e^{a_\al-a_\bet}e^{-\ep_1L_{\al}(s_\al )} e^{\ep_2(A_\bet(s_\al )+1)}+
e^{a_\bet-a_\al} e^{\ep_1(L_{\alpha}(s_\al )+1)} e^{-\ep_2A_\beta(s_\al )} \right]\! ,
\ee
where $L_{\scriptscriptstyle{Y_{\bet}}}(s_\al )$, $A_{\scriptscriptstyle{Y_\bet}}(s_\al )$ are the leg-length  and arm-length factors of the box $s_\al \in Y_\al$ with respect to the Young diagram $Y_\bet$.

For an arbitrary element $v=(\epsilon_1,\epsilon_2,a_{\alpha})$ 
the determinant of $v$ on the tangent space $p_{\scriptscriptstyle{\vec{Y}}}$ obtained from (\ref{char}) reads 
\begin{equation}\label{det-vp}
 \det v_r\Bigl|_{p_{\scriptscriptstyle{\vec{Y}}}}=\prod_{\alpha,\beta=1}^{N}
 \prod_{s_\al \in \scriptscriptstyle{Y_{\alpha}}}
    E_{\scriptscriptstyle{Y_{\alpha}},\scriptscriptstyle{Y_{\beta}}}(a_{\bet}-a_{\al}|s_\al )
   \bigl(\ep-E_{\scriptscriptstyle{Y_{\alpha}},\scriptscriptstyle{Y_{\beta}}}(a_{\bet}-a_{\al}|s_\al )\bigr),
\end{equation}
where $\ep=\epsilon_1+\epsilon_2$ and
\begin{equation}\label{E-def}
    E_{\scriptscriptstyle{Y_{\alpha}},\scriptscriptstyle{Y_{\beta}}}(x|s_\al )=x+\epsilon_{1}(L_{\scriptscriptstyle{Y_{\alpha}}}(s_\al )+1)-\epsilon_{2}\,A_{\scriptscriptstyle{Y_\beta}}(s_\al ).
\end{equation}

Let us now describe the effect of the orbifolding. Even though all fixed points are the same as in the unorbifolded case, the tangent space, which defines the determinants of the vector field, is reduced by the additional constraints \eqref{Z2constr}.  From \eqref{orbaction} we see that the $(i,j)=(1,1)$ box of $Y_\al$ transforms in the $q_\alpha$ representation of $\ZZ_p$; we call $q_{\alpha}\in 0,...,p-1$ the colour of the box. The colour of the boxes in each column of the diagram is the same and differs by $1$ between one column and the next, as in the following example for $\SU(2)$ and $p=2$ with $q_1=0,q_2=1$:
\\[5pt]
\begin{eqnarray}
\label{YT1}
\qquad Y_1=
\begin{picture}(120,0)(0,35)
\multiframe(0,6.5)(15.5,0){1}(15,15){}\put(4,10){$0$}
\multiframe(0,21.5)(15.5,0){2}(15,15){}{}\put(4,25){$0$}\put(20,25){$1$}
\multiframe(0,36.5)(15.5,0){3}(15,15){}{}{} \put(4,41){$0$}\put(20,41){$1$}\put(35,41){$0$}
\multiframe(0,52)(15.5,0){4}(15,15){}{}{}{} \put(4,56){$0$}\put(20,56){$1$}\put(35,56){$0$}\put(51,56){$1$}
\end{picture}
\qquad
Y_2=
\begin{picture}(120,0)(0,40)
\multiframe(0,21.5)(15.5,0){2}(15,15){}{}\put(5,25){$1$}\put(20,25){$0$}
\multiframe(0,36.5)(15.5,0){2}(15,15){}{} \put(5,40){$1$}\put(20,40){$0$}
\multiframe(0,52)(15.5,0){3}(15,15){}{}{} \put(5,55){$1$}\put(20,55){$0$}\put(35,55){$1$}
\end{picture}
\end{eqnarray}
\\\\
In comparison to the ALE case \cite{Fucito:2004b}, we thus obtain striped Young diagrams instead of checkered ones. This is natural since the rows and columns correspond to $\ep_{1,2}$ which correspond to $z_{1,2}$ and since the chain-saw quiver orbifolding only acts on $z_2$ we expect colouring in only one direction. 
While the total number of boxes in the set $\vec{Y}$ is equal to the topological charge (instanton number) $k$, the quantities of cells of different colors determine which connected component of the space $\mathcal{M}_{N,p}(\{k_L\},\{q_\alpha\})$ a given fixed point belongs to. 
More precisely, $k_L$ is equal to the total numbers of boxes with colour $L-1$ (the box with the coordinates $(i,j)$ in the Young diagram $Y_{\alpha}$ has colour $(q_\alpha+j \; \textrm{mod}\,p)$). In the above example $(Y_1,Y_2)\in \mathcal{M}_{2,2}(\{9,8\},\{0,1\})$.

After orbifolding,  
one should keep only the eigenvalues corresponding to the parts of the character that are invariant under \eqref{orbaction}. This implies that 
\begin{equation}\label{det-vp-Zp}
 \det v_{N,p}(\{q_\alpha\})\Bigl|_{p_{\scriptscriptstyle{\vec{Y}}}} \!=\!\prod_{\alpha,\beta=1}^{N} 
 \prod_{s_\al \in \scriptscriptstyle{Y_{\alpha,\beta}^{(0)}}} \!\!\!
    E_{\scriptscriptstyle{Y_{\alpha}},\scriptscriptstyle{Y_{\beta}}}(a_{\bet}{-}a_{\al}|s_\al )\!\!\!
 \prod_{s_\al \in \scriptscriptstyle{Y_{\alpha,\beta}^{(1)}}}  \!\!
   \bigl(\ep{-}E_{\scriptscriptstyle{Y_{\alpha}},\scriptscriptstyle{Y_{\beta}}}(a_{\bet}{-}a_{\al}|s_\al )\bigr),
\end{equation}
where  ($g=0,1$) 
\be\label{regions}
Y_{\alpha,\beta}^{(g)}=\{s_\al \in Y_{\alpha}\, |\, q_\bet-q_\al-A_{Y_\beta}(s_\al )=g \!\! \mod p \}\,.
\ee

Now, we are in a position to formulate the answer for the instanton partition function of the pure  $\mathcal{N}\!=\!2$ $\SU(N)$
gauge theory with a surface defect corresponding to a certain partition specified by a fixed set of $q_{\alpha}$'s. It is given by the following sum over Young diagrams 
\begin{eqnarray} \label{Zgeneral}
\!\! \mathcal{Z}_{N,p}(\{q_{\alpha}\}) \! = \!\sum_{\vec{Y}}  \frac{ \ds \prod_{L=1}^p y_L^{k_L} }{ \ds \prod_{\alpha,\beta=1}^{N} \!
\prod_{s_\al \in \scriptscriptstyle{Y_{\alpha,\beta}^{(0)}}} \!
    E_{\scriptscriptstyle{Y_{\alpha}},\scriptscriptstyle{Y_{\beta}}}(a_{\bet}{-}a_{\al}|s_\al )  \!\!
 \prod_{s_\al \in \scriptscriptstyle{Y_{\alpha,\beta}^{(1)}}}  \!
   \bigl( \ep{-}E_{\scriptscriptstyle{Y_{\alpha}},\scriptscriptstyle{Y_{\beta}}}(a_{\bet}{-}a_{\al}|s_\al )\bigr)} ,
\end{eqnarray}
where $Y_{\alpha,\beta}^{(g)}$ is defined in \eqref{regions}.

As an example, we apply this result to the $\SU(2)$ case with the surface defect specified by $2=1\!+\!1$.  The instanton numbers $k_1,k_2$ are given by
\be \label{Yconstr}
\sum_{j=1}  [ (Y_1)_{2j-1} + (Y_2)_{2j} ] = k_1\,, \qquad  \sum_{j=1} [ (Y_1)_{2j} + (Y_2)_{2j-1} ] = k_2 \,,
\ee
where $Y_j$ denotes the height of column $j$ in $Y$.  The instanton partition function 
for $q_1=0$ (white) and $q_2=1$ (black)  is given by  
\be\label{Z22}
\mathcal{Z}_{2,2}(\{0,1\})  = \sum_{Y_1,Y_2} Z(Y_1,Y_2)  y_1^{k_1} y_2^{k_2} \,,
\ee
where $k_1,k_2$ are the numbers of white and black cells and
\begin{eqnarray}\label{Z22Y1Y2}
&&Z^{-1}(Y_1,Y_2) =\\
&&\prod_{\substack{s_1 \in \scriptscriptstyle{Y_{1}}\\ A(s_1 )-\text{odd}}} \!\!\!
    E_{\scriptscriptstyle{Y_1},\scriptscriptstyle{Y_2}}(-2a|s_1 )(\ep-E_{\scriptscriptstyle{Y_{1}},\scriptscriptstyle{Y_{1}}}(0|s_1 ))
  \!\!      \prod_{\substack{s_1 \in \scriptscriptstyle{Y_{1}}\\ A(s_1 )-\text{even}}} \!\!\!\!
  E_{\scriptscriptstyle{Y_{1}},\scriptscriptstyle{Y_{1}}}(0|s_1 )      
        \bigl(\ep-E_{\scriptscriptstyle{Y_{1}},\scriptscriptstyle{Y_{2}}}(-2a|s_1 )\bigr) \nonumber\\
&&    
\!\! \times \!\!\! \prod_{\substack{s_\al \in \scriptscriptstyle{Y_{2}}\\ A(s_2 )-\text{odd}}} \!\!\!
    E_{\scriptscriptstyle{Y_2},\scriptscriptstyle{Y_1}}(2a|s_2)(\ep-E_{\scriptscriptstyle{Y_{2}},\scriptscriptstyle{Y_{2}}}(0|s_2 ))
 \!\!       \prod_{\substack{s_2 \in \scriptscriptstyle{Y_{2}}\\ A(s_2 )-\text{even}}} \!\!\!\!
    E{\scriptscriptstyle{Y_{2}},\scriptscriptstyle{Y_{2}}}(0|s_2 )
    \bigl(\ep-E_{\scriptscriptstyle{Y_{2}},\scriptscriptstyle{Y_{1}}}(2a|s_2)\bigr) \nonumber .
\end{eqnarray}
Here, according to the definition \eqref{E-def}, the arm-length factors are always calculated with respect to the second diagram in the subscript of the function $E$.

The result (\ref{Zgeneral}) can be related to the expressions in \cite{Feigin:2008,Wyllard:2010b,Kanno:2011} by redefining the variables. For instance, for $\SU(2)$ with the $1\!+\!1$ surface defect one should make the following replacements 
\be \label{newepa}
 \epsilon_2\rightarrow \varepsilon_2/2\,,\qquad \epsilon_1\rightarrow \varepsilon_1\,,\qquad a\rightarrow a+\varepsilon_2/4 \,.  
\ee
Note that choosing for instance $N\!=\!p\!=\!2$ and $q_1\!=\!1$, $q_2\!=\!1$ in (\ref{Zgeneral}), \eqref{regions} leads to a well defined expression, but one that is not directly related to the instanton partition functions in \cite{Feigin:2008,Wyllard:2010b,Kanno:2011}.  It will play a role in the next section.

%%%%%%%%%%%%%%%%%%%%%%%%%%%%%%%%%%%%%%%%%%%%
\section{Combinatorial expressions for  $\cN\!=\!2$ superconformal blocks} \label{sComb}
%%%%%%%%%%%%%%%%%%%%%%%%%%%%%%%%%%%%%%%%%%%%

In section \ref{sRel} we found that the (irregular) $\cN\!=\!2$ superconformal blocks are related to analytically continued (irregular) $\widehat{\sll}(2)$ conformal blocks. More precisely, 
the expression at order $z^N$ in the $\cN\!=\!2$  block is related, up to a factor $M(n)$, to an analytic continuation in $n$ of the expression at order  $x^{n+N} (z/x)^N$ (with $n$ very large) in the $\widehat{\sll}(2)$ block.  The latter expression has a dual gauge theory description \cite{Braverman:2004a,Alday:2010}, arising from the expression at order $y_1^{n+N}y_2^N$ in the instanton partition function of the $\cN\!=\!2$ $\SU(2)$ gauge theory with a surface defect.  This means that the irregular conformal block of the $\cN\!=\!2$ superconformal algebra 
 at order $z^N$ can be obtained from the $\SU(2)$ instanton partition function with a surface defect at order $y_1^{n+N}y_2^N$ via a change of variables, multiplication by $M(n)$, plus analytic continuation in $n$. 
Similarly, the contribution at order $z^{N+\frac{1}{2}}$ in the $\cN\!=\!2$ block can be obtained from the $\SU(2)$ instanton partition function with a surface defect at order $y_1^{n+N+1}y_2^N$ via a change of variables, multiplication by $M(n)$, plus analytic continuation in $n$. 

Below  we present the combinatorial expressions for the irregular $\cN\!=\!2$ superconformal blocks that one obtains by implementing  the analytic continuation in the $\SU(2)$ instanton partition function with a surface defect. We present two different (but equivalent) versions of the result. Details of the second 
derivation are presented in appendix \ref{App1}. Some terms in the expansions for low levels  are collected in appendix \ref{appI}. 

As we saw in section \ref{sSurf},  
the instanton partition function  of the $\cN\!=\!2$ $\SU(2)$ gauge theory with a surface defect  at order $y_1^{n+s+N}y_2^N$ is obtained by summing over fixed points, where the fixed points are labelled by a pair of Young diagrams $(Y_1,Y_2)$ with $k_1 =  n{+}s{+}N $ and $k_2 \!=\! N$ where $k_{1,2}$ were given in \eqref{Yconstr}. The contribution from each fixed point can be determined from the character given in (\ref{usualchar}), (\ref{VW}) subject to the restriction that only terms invariant under (\ref{orbaction}) survive. 
  
Now take $n$ very large. Given the form of $k_{1,2}$ \eqref{Yconstr}, it is easy to see that the first column in the Young
diagram $Y_1$ has to have height $n{+}s{+}r$ where $|r|$ is much smaller than $n$. All other columns in $Y_1$ and $Y_2$ have to have heights smaller than $N\!+\!1$ (otherwise $k_2$ would become too large). Now $Y_1$ without the first column is also a Young diagram, which we denote $\tilde{Y}_1$. We also use the notation $\tilde{Y}_2=Y_2$, as well as 
\be \label{atia}
\tia_1 = a_1-\ep_2\,, \qquad \tia_2 = a_2\,. 
\ee 
It follows that $\tilde{Y}_1$ and $\tilde{Y}_2$ satisfy the constraints 
\be \label{Ytildeconstr}
\sum_{\jmath\ge 1}  [ (\tilde{Y}_1)_{2\jmath-1} + (\tilde{Y}_2)_{2\jmath-1} ] = N\,, \quad  \sum_{\jmath\ge1} [ (\tilde{Y}_1)_{2\jmath} + (\tilde{Y}_2)_{2\jmath} ] = N{-} r \,, \quad 0 \le  r \le N\,,
\ee
where $(\tilde{Y})_\jmath$ denotes the height of column $\jmath$ in $\tilde{Y}$. The restrictions on $r$ follow from the properties of Young diagrams (such as $(\tilde{Y})_\jmath\ge (\tilde{Y})_{\jmath+1}$). 

Now consider the character (\ref{usualchar}). We split $V$ and $V^{*}$ into a piece (a geometric sum) which involves the first column in $Y_1$, plus a piece involving the remaining terms:
\bea \label{Vrel}
V&=&   e^{a_1} \frac{t_1}{t_1-1}(1-e^{-\ep_1(n+s+r)})  + \tilde{V} \,, \non \\
V^*&=&  -e^{-a_1} \frac{1}{t_1-1}(1-e^{\ep_1(n+s+r)}) + \tilde{V}^*\,.
\eea
It is easy to see that $\tilde{V}$ takes exactly the same form as $V$ (\ref{VW}) but with $Y_\al$ and $a_\al$ replaced by $\tilde{Y}_\al$ and $\tilde{a}_\al$. Inserting the  decompositions (\ref{Vrel}) into (\ref{usualchar})  we find 
\be
\chi(a,Y_1,Y_1) = \chi_{\tilde{M}(n)} + \tilde{\chi}(\tilde{a},\tilde{Y}_1,\tilde{Y}_2,r)\,,
\ee
where
\be
\chi_{\tilde{M}(n)} =  (1+e^{\tia_2 - \tia_1})  \sum_{i=1}^{n} e^{i\, \ep_1}\,,
\ee
and 
\bea \label{chinew}
 \tilde{\chi}(\tilde{a},\tilde{Y}_1,\tilde{Y}_2,r) &=& -\tilde{V} \, \tilde{V}^* (t_1-1)(t_2-1)  + \tilde{W}^*\, \tilde{V}+ \tilde{V}^*\,\tilde{W} \, t_1 t_2 \non \\[2pt]
&& +\,  \tilde{U}^*\, \tilde{V} (1-t_2) + \tilde{V}^*\, \tilde{U} (t_1t_2-t_1) +  \tilde{Z}\,,
\eea
with
\be
\tilde{Z}= (1+e^{\tilde{a}_2-\tilde{a}_1}) e^{\ep_1 n} \sum_{i=1}^{r+s} e^{i\, \ep_1} \,, \qquad \tilde{U} = e^{\tilde{a}_1+\ep_2-\ep_1(n+s+r)}\,, \qquad \tilde{W}= \sum_{\al=1}^2 e^{\tilde{a}_\al}\,.
\ee
To obtain the result (\ref{chinew}) we dropped some terms independent of $\tilde{V}$ and $\tilde{V}^*$ that are not invariant under the orbifold projection (\ref{orbaction}) and thus do not contribute. 

Recall that to get the result for the $\cN\!=\!2$ superconformal block at level $N\!+\!\frac{s}{2}$ ($s=0,1$) we should multiply the coefficient at order $x^{n+s+N} (z/x)^N$ in the $\widehat{\sll}(2)$ block by $M(n)$ (\ref{Mn}) and analytically continue in $n$. The factor in the instanton partition function at order $y_1^{n+s+N}y_2^N$ arising from $ \chi_{\tilde{M}(n)}$  simply cancels the $M(n)$ multiplication, provided one uses the map given in \cite{Alday:2010,Kozcaz:2010b} (remembering (\ref{newepa}) and (\ref{atia}))
\be \label{AGT}
\tia_1 = -\tia_2= j +\half \,, \qquad 2 \,\frac{\ep_2}{\ep_1} = -k-2  \,, \qquad y_1=-x \,, \qquad y_2 = -\frac{z}{x}\,.
\ee
In particular, the $(-1)^n$ in $M(n)$ cancels using the relation between $y_1$ and $x$. Thus we are left with the contribution arising from the character (\ref{chinew}). Since in this expression only 
 $\tilde{U}$ and $\tilde{Z}$ depend on $n$ and the dependence is simple, the analytic continuation is trivial, one simply replaces $n$ by $j-\la$.

The result  \eqref{chinew} can be further simplified. Note that the first line in (\ref{chinew}) is simply the character \eqref{usualchar} written in the $\tilde{a}$ variables and can thus be rewritten in the form (\ref{char}). Thus we get a contribution of the same form as in the expression (\ref{Zgeneral}) with (\ref{regions}). However, there is an important difference: Since the original Young diagrams $Y_1$ and $Y_2$ had $q_1\!=\!0$ and $q_2\!=\!1$, the $\tilde{Y}_\al$ will have $q_1\!=\!1$ and $q_2\!=\!1$. Apart from this difference, the first line in (\ref{chinew}) gives the same contribution as for the $\SU(2)$ theory with a surface defect and $k_1=N$, $k_2=N{-}r$. The contribution from the terms on the second line of \eqref{chinew} are easily determined.

We will use the notation $\lb W|W\rb^{(s)}$ where  $s=0,1$ for the norm of the Whittaker state (i.e.~the irregular block). In the notation of sections \ref{sN2} and \ref{ssl2} we have $\lb W|W\rb^{(0)}\equiv \lb W|W\rb$ and $\lb W|W\rb^{(1)}\equiv{}^-\lb W|W\rb^{-}$.  The combinatorial  result for $\lb W|W\rb^{(s)}$ follows by putting everything together 
\bea \label{endresult1}
&&\lb W|W\rb^{(s)}   = \sum_{N=0}^{\infty}  \sum_{\tY_1,\tY_2, r} z^{ (|\tY_1|+|\tY_2|+s+r)/2} \,\, \frac{1}{\ep_1^{2(r+s)} (-1)^{r+s} (n {+} 1)_{r+s}(-\frac{2\tia}{\ep_1}{+}n{+}1)_{r+s}} \non \\    
&& \!\! \!\! \!\!  \times \prod_{\alpha=1}^{2} 
\prod_{s_\al\in \tY^{(0)}_{\al} } \!\!
\frac{F(\tia_\al{-}\tia_1{-}\ep_2{+}\ep_1(n{+}r{+}s{-}1) |s_\al)}{F(\tia_\al{-}\tia_1{-}\ep_2{+}\ep_1(n{+}r{+}s) |s_\al)}  \!
 \prod_{s_\al\in \tY^{(1)}_{\al} }  \!\!
 \frac{F(\tia_\al{-}\tia_1{+}\ep_1(n{+}r{+}s)  |s_\al)  }
 {F(\tia_\al{-}\tia_1{-}2\ep_2{+}\ep_1(n{+}r{+}s{-}1)|s_\al)}
\non \\ 
 &&  \times\bigg[\prod_{\alpha,\beta=1}^{2}\prod_{s_\al\in \scriptscriptstyle{\tY_{\alpha,\beta}^{(0)}}}
    E_{\scriptscriptstyle{\tY_{\alpha}},\scriptscriptstyle{\tY_{\beta}}}(\tia_{\bet}-\tia_{\al}|s_\al)  
 \prod_{s_\al\in \scriptscriptstyle{\tY_{\alpha,\beta}^{(1)}}}  
   \bigl(\ep_1+\ep_2{-}E_{\scriptscriptstyle{\tY_{\alpha}},\scriptscriptstyle{\tY_{\beta}}}(\tia_{\beta}-\tia_{\alpha}|s_\al)\bigr)\bigg]^{-1} \!\!,
\eea
where the sum over $(r,\tY_1,\tY_2)$ is constrained by the relations (\ref{Ytildeconstr}),
 \bea
F (x|s_\al) \! \!&=&\! \! x  -\ep_1 (i_\al-1)- \ep_2 (j_\al-1) \,, \non \\
 E_{\scriptscriptstyle{\tY_{\alpha}},\scriptscriptstyle{\tY_{\beta}}}(x|s_\al) \!\!&=&\! \! x+\ep_1(L_{\scriptscriptstyle{\tY_{\alpha}}}(s_\al)+1)-\ep_2\,A_{\scriptscriptstyle{\tY_\beta}}(s_\al)\,,
\eea
and
\bea
&&\tY_{\alpha}^{(g)}=\{s_\al=(i_\al,j_\al)\in \tY_{\alpha}\, |\, j_\al=g\!\! \mod 2 \} \,, \non \\
&&\tY_{\alpha,\beta}^{(g)}=\{s_\al\in \tY_{\alpha}\, |\, A_{\tY_\beta}(s_\al)=-g\!\! \mod 2 \} \,.
\eea
After using the relations 
\be \label{AGT2}
n=j-\la\,, \qquad \tia_1 = -\tia_2= j +\half \,, \qquad 2 \, \frac{\ep_2}{\ep_1} = -k-2 \,,
\ee
our result (\ref{endresult1}) gives the irregular $\cN\!=\!2$ superconformal blocks written in terms of relaxed $\widehat{\sll}(2)$ variables. To translate to $\cN\!=\!2$ variables one uses (\ref{dj}), (\ref{ck}) and (\ref{zztrans}). 

Note that  for the relaxed $\widehat{\sll}(2)$  case the above expression (\ref{endresult1}) also makes sense for any non-negative integer $s$, so that we have in fact computed the generating function
\be
Z_{\widehat{\sll}(2)\mathrm{rel}}(x,z) = \sum_{s=0}^{\infty}  \lb W|W\rb^{(s)}  x^s  \,.
\ee

%%%%%%%%%%%%%%%%%%%%%%%%%%%%%%%%%%%%%%%%%%%%
\subsection{Alternative combinatorial expression for  $\cN\!=\!2$ superconformal blocks} \label{salt}
%%%%%%%%%%%%%%%%%%%%%%%%%%%%%%%%%%%%%%%%%%%%

Below we present an alternative combinatorial expression for the irregular $\cN\!=\!2$ superconformal blocks at level $N{+}\frac{s}{2}$ where $s=0,1$ (corresponding to level $N$, charge $s$ in the relaxed $\widehat{\sll}(2)$ block). This expression involves a sum over pairs of Young diagrams $(Y_1,Y_2)$ satisfying the constraints  $k_1\!=\!2k_2\!=\!2 N$  with $k_{1,2}$ as in (\ref{Yconstr}). 
  The number of such fixed points can be shown to be equal to the number of states predicted by the coset (1.2), cf.~\eqref{chiN2} and \eqref{chiFix} (recall that the coset also contains a decoupled free boson). 

We will use the notation $\lb N|N\rb^{(s)}$ for the norm of the Gaiotto states. In the notation of sections \ref{sN2} and \ref{ssl2} we have $\lb N|N\rb^{(0)}\equiv \lb N|N\rb= \lb N,0 | N,0\rb$ and $\lb N|N\rb^{(1)}\equiv{}^-\lb N|N\rb^{-}= \lb N,-|N,-\rb$.  
The details of the derivation of the alternative combinatorial  result for $\lb N|N\rb^{(s)}$ are collected in appendix \ref{App1}. 
\begin{eqnarray}\label{N2comb}
\langle N|N\rangle^{(s)} = \Om
 \sum_{\substack{Y_1,Y_2\\\#\Box= 2N,\\\#\blacksquare = N}} K(Y_1,Y_2)^{-1}\,,
\end{eqnarray}
where  
\begin{equation} \label{Omega}
\Om=\frac{\prod_{m=1}^{N-s} [\epsilon_1(n{-}m{+}1)][-2\tia{+}\epsilon_1(n{-}m{+}1)] }{(-1)^s \prod_{m=1}^{s-N} [\epsilon_1(n{+}m)][-2\tia{+}\epsilon_1(n{+}m)] }\,,
\end{equation}
and
\bea
\label{FY1Y2}
&&  \!\!\!  \!\!\!  \!\!\! K(Y_1,Y_2) =  \\
&&   \!\!\! \!\!\! \!\!\!   
  \!\!\!   \prod_{\substack{s_1\in \scriptscriptstyle{Y_{1}}\\ A(s_1)-\text{odd}}}  \!\!\!
   \!\!   \hat{E}_{\scriptscriptstyle{Y_1},\scriptscriptstyle{Y_2}}(-2\tia{-}\ep_2|s_1)
   (\ep- \hat{E}_{\scriptscriptstyle{Y_{1}},\scriptscriptstyle{Y_{1}}}(0|s_1))
  \!\!   \!\!    \!\!\!     \prod_{\substack{s_1\in \scriptscriptstyle{Y_{1}}\\ A(s_1)-\text{even}}}  \!\!\!  \!\! 
 \!\!   \hat{E}_{\scriptscriptstyle{Y_{1}},\scriptscriptstyle{Y_{1}}}(0|s_1)      
        \bigl(\ep-\hat{E}_{\scriptscriptstyle{Y_{1}},\scriptscriptstyle{Y_{2}}}(-2\tia{-}\ep_2|s_1)\bigr) \nonumber\\
   &&   \!\!\!  \!\!\! \!\!\! \!\!  \times  \!
  \!\!\!  \!\! \prod_{\substack{s_2\in \scriptscriptstyle{Y_{2}}\\ A(s_2)-\text{odd}}}  \!\!\!  \!\! 
    E_{\scriptscriptstyle{Y_2},\scriptscriptstyle{Y_1}}(2\tia{+}\ep_2|s_2)(\ep-E_{\scriptscriptstyle{Y_{2}},\scriptscriptstyle{Y_{2}}}(0|s_2))
 \!\!\!    \!\!  \!\!\!   \prod_{\substack{s_2\in \scriptscriptstyle{Y_{2}}\\ A(s_2)-\text{even}}}  \!\!\!  \!\! 
    E{\scriptscriptstyle{Y_{2}},\scriptscriptstyle{Y_{2}}}(0|s_2)
    \bigl(\ep-E_{\scriptscriptstyle{Y_{2}},\scriptscriptstyle{Y_{1}}}(2\tia{+}\ep_2|s_2)\bigr)  , \non
\eea
where $E_{\scriptscriptstyle{Y_{\alpha}},\scriptscriptstyle{Y_{\beta}}}(x|s_\al)$ is as previously defined  \eqref{E-def} and
\be\label{tildeE-def}
    \hat{E}_{\scriptscriptstyle{Y_{\alpha}},\scriptscriptstyle{Y_{\beta}}}(x|s_\al)=
    \begin{cases}
    x+\epsilon_1(L_{Y_{\alpha}}(s_\al){+}n{+}s{-}N{+}1)-\epsilon_2 A_{Y_\beta}(s_\al),\quad s_1 \in \text{first column}\\
   x+\epsilon_1(L_{Y_{\alpha}}(s_\al)+1)-\epsilon_2 A_{Y_\beta}(s_\al),\qquad\qquad  \qquad \;\; \;\,\text{otherwise}
    \end{cases}
\ee
Using (\ref{AGT2}), the result (\ref{N2comb}) gives the norm of the $\cN\!=\!2$ Gaiotto states written in terms of relaxed $\widehat{\sll}(2)$ variables. To translate to $\cN\!=\!2$ variables one uses (\ref{dj}), (\ref{ck}) and (\ref{zztrans}). The seemingly different looking results (\ref{endresult1}) and (\ref{N2comb}) are equivalent as is clear from how they were derived.

 We close this section by stressing that since the relation between the instanton partition function in the $\cN\!=\!2$ $\SU(2)$ theory with a surface defect and the 
unrelaxed irregular $\widehat{\sll}_2$ blocks was proved in \cite{Braverman:2004a}  and since we have shown the relation between the irregular $\cN=2$ and the relaxed $\widehat{\sll}_2$ blocks and related them to the unrelaxed irregular $\widehat{\sll}_2$ blocks, we have a complete derivation of the combinatorial expressions presented above.

%%%%%%%%%%%%%%%%%%%%%%%%%%%%%%%%%%%%%%%%%%%%
\subsection{String theory interpretation of the analytic continuation} \label{string}
%%%%%%%%%%%%%%%%%%%%%%%%%%%%%%%%%%%%%%%%%%%%

The analytic continuation argument presented above led to a completely explicit combinatorial expression for the $\cN\!=\!2$ superconformal blocks in the Gaiotto limit. However, the gauge theory and instanton moduli space interpretation of the result is not very clear. 

In this subsection we give an outline of a string theory interpretation of the analytic continuation in the hope that this argument will lead to a deeper understanding and facilitate future generalisations. Our discussion is not intended to be rigorous and several aspects have not been worked out in detail. 

 It is known \cite{Witten:1995b} that the ADHM construction in the $\SU(N)$ gauge theory can be  realised in type II string theory  by considering the world-volume theory on $k$ D(-1)-branes on top of $N$ (coincident) D3-branes. Here $k$ is the instanton number and the world volume theory contains in particular the $B_{1,2}$ matrices which geometrically are associated with the two complex coordinates $z_{1,2}$ of the D3-brane world-volume. 
 
The $\ZZ_2$ orbifold splits the $k$  D(-1)-branes into two sets containing $k_1$ and $k_2$  D(-1)-branes each, and projects out some of the modes in the world-volume theory on the D(-1)-branes.  As we saw above, the two off-diagonal blocks in $B_1$ are projected out leaving the two diagonal blocks, whereas the two diagonal blocks in $B_2$ are projected out leaving the two off-diagonal blocks. 

A peculiar feature of the non-abelian nature of D-branes occurs when one takes the large $n$ limit of a stack of $n$ D(p)-branes in a particular string theory background with non-zero (constant) RR fluxes. It is known that in the large $n$ limit the stack of D$(p)$-branes can equivalently be viewed as a {\it single} D$(p{+}2)$-brane. This is known as the Myers (dielectric) effect \cite{Myers:1999} (see  e.g.~\cite{Myers:2003} for a review). 

In our case we have $k_1 \!=\! n\!+\!N$ where we take $n$ to be large so the arguments in \cite{Myers:1999} potentially apply in our example. The fact that we have non-zero $\ep_{1,2}$ parameters means that the gauge theory is in the so called $\Omega$-background, which can alternatively be interpreted as a background with non-trivial RR fluxes \cite{Billo:2006}. These RR fluxes have precisely the right form for the arguments in \cite{Myers:1999} to apply. Finally, since it is only $B_1$ that has  a $k_1{\times}k_1$ matrix component, the polarisation into a higher-dimensional D-brane can only involve the $z_1$ direction. Thus we conclude that in the large $n$ limit, $n$ D(-1)-branes expand into an (spherical) euclidean D(1)-brane intersecting the surface defect (which is located at $z_2=0$). The parameter $n$ is related to the flux of the $\U(1)$ gauge field on the D(1)-brane. 

There should be different sectors in which $r$ of the remaining $N$ D(-1)-branes in the first stack are bound to the D1-brane. The remaining $2N-r$ D(-1)-branes then have $k_1= N{-}r$ and $k_2 = N$. These should give a contribution to the instanton partitin function, which is the $\ZZ_2$ orbifold contribution, except that the $q_\al$ (which have an interpretation in terms of holonomies, see e.g.~\cite{Fucito:2004b,Ito:2011}) have changed due to the presence of the D(1)-brane.  In addition, there are also degrees of freedom coming from the strings connecting the  $2m{-}r$ D(-1)-branes to the D1-brane, as well as a contribution from the $r$ D(-1) branes bound to the D(1)-brane.  This gives a qualitative explanation for the form of the instanton character (\ref{chinew}). 
 It should be possible to make this argument more precise, perhaps using arguments as in \cite{Matsuura:2008}. From the point of view of the gauge theory, it seems that the end result can be viewed as a surface defect with additional structure.  For a discussion of similar objects in a related context see e.g.~\cite{Gaiotto:2011} and references therein.

%%%%%%%%%%%%%%%%%%%%%%%%%%%%%%%%%%%%%
\section{Concluding remarks}\label{Concl}
%%%%%%%%%%%%%%%%%%%%%%%%%%%%%%%%%%%%%

In this paper we studied  $2d$ $\cN\!=\!2$  superconformal field theories in the context of the correspondence  between $2d$ CFTs and $4d$ $\cN\!=\!2$ gauge theories. 

One of the basic conceptual questions is how to find the CFT symmetry algebra associated with a given instanton moduli space and vice versa. Once the answer for a specific example is known, one may deduce many useful consequences. One outcome  is the relation  \cite{Alday:2009a} between conformal blocks and the instanton partition functions (which are computable by summing over fixed points). Another important by-product is the integrable structure of the CFT which becomes explicit in a special basis \cite{Alba:2010}.

Even though there are many examples where dual pairs have been found, 
there is no canonical or ``direct'' way to pinpoint the CFT which is relevant
for a particular instanton moduli space. One suggestive observation is that many  symmetry algebras in $2d$ CFTs arise as a special limit  of so called toroidal  algebras.  Toroidal algebras also appear in the gauge theory context since they have a natural action on the equivariant homologies of  instanton moduli spaces.  
Remarkably, the symmetry algebras one obtains from toroidal algebras via this limit  
usually involve various cosets, suggesting that there is a way to obtain CFT symmetry algebras  realised as a cosets from instanton moduli spaces.

In this paper we have in a sense solved the inverse problem since we started from the coset realisation of the $\cN\!=\!2$ superconformal algebra in terms of the $\widehat{\sll}(2)$ algebra. 
 We studied the relation between the highest weight representations of the $\cN\!=\!2$ algebra
and the rather unusual relaxed representations of $\widehat{\sll}(2)$. For the
regular $\widehat{\sll}(2)$ case the gauge theory dual involves the instanton moduli space, $\mathcal{M}_{2,2}$, corresponding to the $\cN\!=\!2$ $\SU(2)$ gauge theory with a surface defect.
We studied the consequences of the relaxation procedure from the point
of view of this dual moduli space description. We found that the relaxation procedure leads to a  another moduli space, $\mathcal{M}^{\star}_{2,2}$, which is closely related to  $\cM_{2,2}$. 
The fixed points of the vector field associated with $\mathcal{M}^{\star}_{2,2}$  satisfy an interesting stabilisation condition that we described in appendix \ref{App1}. We found that the number of fixed points  in $\mathcal{M}^{\star}_{2,2}$ coincides with the dimensions of the representation space of the coset realisation of the  $\cN\!=\!2$ algebra. 

This lead us to propose that the moduli space $\mathcal{M}^{\star}_{2,2}$
is related to the $\cN\!=\!2$ superconformal field theory. We checked this idea 
for the simplest example, viz.~the four-point conformal block function in the Gaiotto limit.
We showed that this (irregular) conformal block is equal to the instanton partition function associated with $\mathcal{M}^{\star}_{2,2}$. This instanton partition function  is related to an analytically continued version of the instanton partition function for the pure $\cN\!=\!2$ $\SU(2)$ gauge theory with a surface defect, but the gauge theory interpretation of the final result is not completely clear. In particular, one would like to understand in more detail the transition the moduli space undergoes when one implements the analytic continuation. The string theory interpretation presented in section \ref{string} should prove to be useful. 

Our main result for the irregular conformal blocks was presented in two versions and is contained in (\ref{endresult1}) and  (\ref{N2comb}).
These combinatorial expressions  involve sums over pairs of two-coloured striped Young diagrams (subject to certain restrictions). This should be compared to the $\cN\!=\!0$ blocks which involve a sum over a pair of uncoloured Young diagrams \cite{Alday:2009a}, and the $\cN\!=\!1$ blocks which involve a sum over  a pair of two-coloured checkered Young diagrams \cite{Belavin:2011a}.  

The instanton partition function for the $\cN\!=\!2$ $\SU(2)$ gauge theory with a surface defect also has a dual description in terms of a degenerate state insertion in the Liouville theory  \cite{Alday:2009b}, which in turn can be viewed as an $\SU(2){\times}\SU(2)$ theory with a certain restriction on the parameters of the theory. This alternative description is closely related to topological strings and toric branes  (see e.g.~\cite{Kozcaz:2010} for more details). It would be interesting to investigate if it is possible to implement and interpret  the analytic continuation in this language. 

There are several  other possible extensions of our results. A natural extension is to consider proper conformal blocks. It should also be possible to consider intermediate cases along the lines of \cite{Kanno:2012}. Another
interesting problem is to consider the full CFT correlation functions and try to find a gauge theory interpretation as was done for ($\cN\!=\!0$) Liouville in  \cite{Alday:2009a}  and for $\cN\!=\!(1,1)$ super-Liouville in  \cite{Bonelli:2011b}. The natural candidate for the full CFT in the $\cN\!=\!2$ case is the $\cN\!=\!(2,2)$ super-Liouville theory (see e.g.~\cite{Hosomichi:2004} and references therein). There is a whole class of coset theories with $\cN\!=\!2$ supersymmetry, the so called Kazama-Suzuki models \cite{Kazama:1988}. Is is possible to compute conformal blocks also for such models? It would also be interesting to see if one can understand our combinatorial formul\ae{} directly within the CFT via a suitable choice of basis, as was done for $\cN\!=\!0$ (Virasoro) in  \cite{Alba:2010,Belavin:2011}  and for $\cN\!=\!1$ (RNS) in \cite{Belavin:2011c}. We hope to return to some of these questions in the near future.

%%%%%%%%%%%%%%%%%%%%%%%%%%%%%%%%
\section*{Acknowledgements}
%%%%%%%%%%%%%%%%%%%%%%%%%%%%%%%%

The initial discussions which ultimately led to this paper took place during the workshop ``Geometric Correspondences of Gauge Theories'' held at SISSA in September 2011.
We are grateful to Giulio Bonelli, Kazunobu Maruyoshi and Alessandro Tanzini 
for the invitations. 

V.B. thanks Alexander Belavin, Samuel Belliard and Boris Feigin for many 
useful discussions. He also thanks the organizers of the Nordita program on Exact 
Results in Gauge-String Dualities for the hospitality and stimulating scientific 
atmosphere. 
He is grateful to the members of the L2C laboratory  at the University II of Montpellier  
and especially to Andre Neveu for the hospitality while this work was in progress. 
His work was supported by RFBR No.12-01-00836-a.  

N.W. would like to thank ICTP-SAIFR in S\~ao Paulo for generous hospitality while part of this work was in progress. Part of the work was also done while he was attending the workshops ``New perspectives in supersymmetric gauge theories" at the ASC in Munich, and ``Workshop on the AGT conjecture" at the BCTP in Bonn. He would like to thank the organisers for invitations to these inspiring meetings.  Finally he would like to thank Anatoly Konechny and Richard Szabo for discussions and hospitality at Heriot-Watt University during the final stages of this project.

%%%%%%%%%%%%%%%%%%%%%%%%%%%%%%%%%%%%%%%%%%%%%%%%%%%%%%%%%%%%%%%%%%%%%%%%%%%%%%%%
\appendix
\section*{Appendix}
%%%%%%%%%%%%%%%%%%%%%%%%%%%%%%%%%%%%%%%%%%%%%%%%%%%%%%%%%%%%%%%%%%%%%%%%%%%%%%%%

%%%%%%%%%%%%%%%%%%%%%%%%%%%%%%%%%%%%%%%%%%%%
\section{Derivation of alternative combinatorial expression} \label{App1}
%%%%%%%%%%%%%%%%%%%%%%%%%%%%%%%%%%%%%%%%%%%%

Here we discuss an alternative way to implement the analytic continuation defined in section \ref{sanal} in the  instanton partition function for the $\SU(2)$ theory with a surface defect.

As discussed in section \ref{sanal} the relaxed and unrelaxed  
$\widehat{\sll}(2)$  conformal blocks are related in the following way
\begin{eqnarray} \label{NNM0}
\langle N|N\rangle^{(s)}_{\text{sl2rel}}=\big[M(n)\langle N|N\rangle^{(s+n)}_{\text{sl2}}\big]
\big{|}_{n\rightarrow j-\lambda}\,.
\end{eqnarray}
For $\langle N|N\rangle_{\text{sl2}}$ we will use the representation 
in terms of the instanton partition function (\ref{Z22}), (\ref{Z22Y1Y2}) with $q_1=0,q_2=1$
\begin{equation}\label{NN}
\langle N|N\rangle^{(s+n)}_{\text{sl2}}= (-1)^{n+s} \sum_{Y_1,Y_2} [Z_{Y_1,Y_2}]^{-1}.
\end{equation}
From the results in section \ref{sSurf}, it follows that the sum goes over 
Young diagrams with $N{+}s{+}n$ white boxes  and  $N$  black boxes.
Since we are interested in the analytic continuation, we may assume that 
$n$ is big enough to make $N{+}s{+}n> 2N$. The factor $(-1)^{n+s}$ arises from the relation 
$x\!=\!-y_1$ between the expansion parameters in the two descriptions.

In order for the analytic continuation to make sense it is crucial that the number of pairs of Young diagrams in \eqref{NN} does not depend on $n$. To show this we compute the character of the relevant pairs of Young diagrams. 
  
We consider first the case of one diagram. The generating function is
\begin{equation}\label{char1}
\chi^{(1)}(q_1,q_2)\equiv\sum_{k_1,k_2} N^{(1)}(k_1,k_2) q_1^{k_1} q_2^{k_2},
\end{equation}
where $N(k_1,k_2)$ is the number of diagrams with $k_1$ white boxes and
$k_2$ black boxes.
It has been shown in \cite{Canfield:2003}  that for $k_1\ge 2 k_2$, the number of diagrams 
$N^{(1)}(k_1,k_2)$ only depends on $k_2$  and is related to the boson character (\ref{chibos})
as follows:
\begin{eqnarray}
N^{(1)}(k_1,k_2)=N_b^{(2)}(k_2)\,, \quad\text{for}\quad k_1\ge 2k_2\,
\end{eqnarray}
where $N_b^{(2)}(k)$ is defined as 
\begin{equation}
[\chi_{\text{boson}}(q)]^2=\sum_{k} N_b^{(2)}(k) q^{k}.
\end{equation}
Similarly, for the generating function of pairs of Young diagrams with $k_1$ white and $k_2$ black 
cells, 
\begin{equation}\label{char2}
\chi^{(2)}(q_1,q_2)\equiv\sum_{k_1,k_2} N^{(2)}(k_1,k_2) q_1^{k_1} q_2^{k_2},
\end{equation}
provided that $k_1\ge 2k_2$, we find 
\begin{eqnarray}
N^{(2)}(k_1,k_2)=N_b^{(4)}(k_1)\,, \quad\text{for}\quad k_1\ge 2k_2\,,
\end{eqnarray}
and
\begin{equation}
[\chi_{\mathrm{boson}}(q)]^4=\sum_{k} N_b^{(4)}(k) q^{k}\,.
\end{equation}
Thus,  the number of fixed points, relevant in the correspondence between instanton counting and
the $\cN=2$ superconformal theory, is given by 
\begin{equation} \label{chiFix}
[\chi_{\mathrm{boson}}(q)]^4=1+4 q+14q^2+40q^3+105 q^4+...
\end{equation}

From the fact that the number of pairs of Young diagrams is independent of $n$ and is the same as for the number of white boxes equal to $2N$ it follows that
\begin{equation}\label{NN1}
\langle N|N\rangle^{(s+n)}_{\text{sl2}}= (-1)^{n+s} \!\!\!\!\!  \!\!\!\! \!\!  \sum_{\substack{Y'_1,Y_2\\\#\Box= N+s+n
,\#\blacksquare=N }}   \!\! \!\! \!\! [Z_{Y'_1,Y_2}]^{-1}= \!\!\!\!  \!\!\!\!
\sum_{\substack{Y_1,Y_2\\\#\Box= 2N,\#\blacksquare = N}} [\hat{Z}_{Y_1,Y_2}]^{-1},
\end{equation}
where the only difference between the Young diagrams $Y'_1$ and $Y_1$ is that  $Y'_1$ contains $n{+}N{+}s{-}2N=n{+}s{-}N$ additional boxes in the first column.  We used the notation $\hat{Z}$ to stress that the expression arises 
from $Z$ in the first sum and is not the standard expression.
Thus, the problem is reduced to the evaluation of the ratio
\begin{equation}\label{Zsplit}
\frac{(-1)^{n+s} M(n)}{Z_{Y'_1,Y_2}}
=
\frac{(-1)^s Z_{1^n,\varnothing}} {Z_{Y'_1, Y_2}\big|_{s_1 =(m,1) }} 
\, \bigl[ Z_{Y_1,Y_2}\bigr]^{-1}_{ s_1 \ne(m,1) } \,,
\end{equation}
where we have separated out the contribution from boxes in the first column of $Y'_1$, and  
used that $M(n)= (-1)^n Z_{\varnothing,1^n}$. The contribution from the boxes outside the first column of $Y'_1$ does not depend on $n$ and is the same as the contribution with $Y'_1$ replaced by $Y_1$.

Let us compute the contribution from the first column. For the boxes in the first column of $Y'_1$ that also belong to $Y_1$ the contribution is the standard one except that the leg length $L_{Y_1}(s_1)$ is replaced by $L_{Y_1}(s_1) + n{+}s{-}N$. In other words, when $s_1$ belongs to the first column of $Y_1$ we get the standard contribution but with $E_{Y_1,Y_\bet}$ replaced by $\hat{E}_{Y_1,Y_\bet}$ defined in (\ref{tildeE-def}). 

The boxes  in the first column of $Y'_1$ that do not belong to $Y_1$ by construction do not have any boxes to the right of them,  i.e.~have zero arm-length with respect to $Y'_1$ and arm-length $-1$ with respect to $Y_2$. Furthermore, the leg length of box $r$ is $n{+}s{-}N{-}r$. Thus they give rise to the contribution 
\be \label{DOWN}
\prod_{r=1}^{n+s-N}  [-2a+\epsilon_1(n+{s}{-}N{-}r+1)+\epsilon_2]  [\epsilon_1(n+{s}{-}N{-}r+1)]\,.
\ee
On the other hand, we have
\be \label{UP}
Z_{1^n,\varnothing}  =  \prod_{r=1}^{n}  [-2a+\epsilon_1(n-r+1)+\epsilon_2]  [\epsilon_1(n-r+1)] \,.
\ee
If $s\!=\!N$ the ratio of (\ref{UP}) and (\ref{DOWN})  is 1. If $s>N$ we get a residual contribution in the numerator, whereas  if $s<N$ get a residual contribution in the denominator. These two contributions (plus the $(-1)^s$ factor) give the contribution (\ref{Omega}). 

Combining all the pieces,  we finally arrive at the result \eqref{N2comb}. We stress that this expression uses $q_1=0$ and $q_2=1$. Also $2\tia = 2a-\ep_2$.

%%%%%%%%%%%%%%%%%%%%%%%%%%%%%%%%
\section{Examples for low instanton numbers} \label{appI}
%%%%%%%%%%%%%%%%%%%%%%%%%%%%%%%%
In this appendix we list the explicit expressions the first few terms in the combinatorial expressions for the irregular $\cN\!=\!2$ blocks given in \eqref{endresult1} or equivalently in \eqref{N2comb}.
These two alternative expressions use different ways to label the fixed points. 
In version I the fixed points are labelled by two Young diagrams and an integer $r$ subject to the restrictions (\ref{Ytildeconstr}). In version II the fixed points are labelled by a pair of Young diagrams with $2N$ white boxes and $N$ black boxes. It is easy to show that the number of fixed points is the same in the two cases and to find the one-to-one map. Consider version II. Since $k_2=N$ the number of boxes in the first column of $Y_1$ has to be at least $N$ (the sum of the remaining columns that contribute to $k_1$ are bounded above by $k_2$ by the properties of Young diagrams). This means that one finds different classes where in a given class the first column of $Y_1$ contains $N{+}r$ boxes with $0\le r\le N$. Removing the first column of $Y_1$  the remaining pieces form two Young diagrams of precisely the same type as in version I, which shows that the number of fixed points is the same.

At level 1, using the version I language, the fixed points are labelled by
\be
(r,\tY_1,\tY_2)=(0,\tableau{2},\varnothing)\,, \quad (0,\varnothing,\tableau{2})\,, \quad   (1,\tableau{1},\varnothing) \,,  \quad  \mathrm{and} \quad(1,\varnothing,\tableau{1})\,.
\ee
whereas using the version II language the fixed points are labelled by
\be
(Y_1, Y_2)=(\tableau{1},\tableau{2})\,, \quad (\tableau{2 1},\varnothing)\,, \quad (\tableau{1 1},\tableau{1})\,,  \quad \mathrm{and} \quad  (\tableau{3},\varnothing)\,.
\ee
At level 0 there is a contribution for $s=1$ given by
\be
-\frac{1}{\ep_1 (1 {+} n) (-2 \tia {+} \ep_1 {+} \ep_1 n)}\,.
\ee
At level 1, the $s=0$ contribution is
\bea 
&&-\, \frac{n}{4\, \tia \,\ep_2 (-2 \tia {+} \ep_1 {+} 2 \ep_2) (2 \ep_2 {-} \ep_1 n)} 
- \frac{1}{ 2 \, \tia \, \ep_1 (-2 \tia {+} \ep_1 {+} \ep_1 n) (-2 \ep_2 {+} \ep_1 n)} \\
&& +\,  \frac{1}{ 2\, \tia \, \ep_1^2 (1 + n) (-2 \tia - 2 \ep_2 + \ep_1 n)} 
- \frac{-2 \tia + \ep_1 n}{ 4\, \tia \, \ep_1 \, \ep_2 (2 \tia + \ep_1 + 2 \ep_2) (-2 \tia - 2 \ep_2 + \ep_1 n)}\,, \non
\eea
whereas for $s=1$ one finds instead
\bea
&&-\,\frac{1}{4 \, \tia \, \ep_1 \, \ep_2 (-2 \tia {+} \ep_1 {+} 2 \ep_2) (-2 \tia {+} \ep_1 {+} \ep_1 n) (\ep_1 - 2 \ep_2 {+}  \ep_1 n)} \non \\
     &&+\, \frac{1}{ 2\, \tia \, \ep_1^2 (1 {+} n) (-2 \tia {+} \ep_1 {+} \ep_1 n) (-2 \tia {+} 2 \ep_1 {+} \ep_1 n) (\ep_1 {-} 2 \ep_2 {+}
     \ep_1 n)}  \non \\
     &&+ \,\frac{1}{ 4 \, \tia \, \ep_1^2\, \ep_2 (2 \tia + \ep_1 {+} 2 \ep_2) (1{ +} n) (-2 \tia {+} \ep_1 {-} 2 \ep_2 + 
    \ep_1 n)}  \\
    &&- \,\frac{1}{ 2 \, \tia \, \ep_1^3 (1 {+} n) (2 {+} n) (-2 \tia {+} \ep_1 {+} \ep_1 n) (-2 \tia {+} \ep_1 {-} 2 \ep_2 {+} \ep_1 n)} \non\,.
\eea

\providecommand{\href}[2]{#2}\begingroup\raggedright\endgroup

\end{document}